\documentclass[a4paper,fleqn,usenatbib]{mnras}

\usepackage{newtxtext,newtxmath}

\usepackage[T1]{fontenc}
\usepackage{ae,aecompl}

\usepackage{graphicx}	
\usepackage{amsmath}	
\usepackage{amssymb}	

\usepackage{bm}

\newcommand{\V}{\bm{v}}

\newcommand{\rp}{r_{\rm p}}

\newcommand{\Mp}{M_{\rm p}}

\newcommand{\au}{{\, \rm au}}

\newcommand{\xs}{x_{\rm s}}
\newcommand{\cs}{c_{\rm s}}
\newcommand{\alphaSS}{\alpha_{\rm ss}}

\hypersetup{draft}   

\title[Super-Earths in low-viscosity discs]{Migrating super-Earths in low-viscosity discs: unveiling the roles of feedback, vortices, and laminar accretion flows}

\author[C.~P.~McNally et al.]{
Colin P.~McNally,$^{1}$\thanks{E-mail: colin@colinmcnally.ca (CPM)}
Richard P.~Nelson, $^{1}$ Sijme-Jan Paardekooper $^{1,2}$ \newauthor
and  Pablo Ben\'itez-Llambay $^{3}$
\\
$^{1}$Astronomy Unit, School of Physics and Astronomy, Queen Mary University of London, London E1 4NS, UK\\
$^{2}$DAMTP, University of Cambridge, Wilberforce Road, Cambridge CB3 0WA, UK\\
$^{3}$Niels Bohr International Academy, Niels Bohr Institute,
Blegdamsvej 17, DK-2100 Copenhagen \O, Denmark
}

\date{Accepted: 2018 December 28; Revised: 2018 December 14; Received: 2018 October 10}

\pubyear{2018}

\begin{document}
\label{firstpage}
\pagerange{\pageref{firstpage}--\pageref{lastpage}}
\maketitle


\begin{abstract}
We present the highest resolution study to date of super-Earths migrating in inviscid and low-viscosity discs, motivated by the connection to laminar, wind-driven models of protoplanetary discs. 
Our models unveil the critical role of vortices in determining the migration behaviour for partial gap-opening planets. Vortices form in pressure maxima at gap edges, and prevent the disc-feedback stopping of migration for intermediate planets in low-viscosity and inviscid discs, contrary to the concept of the `inertial limit' or  `disc feedback' halting predicted from analytical models. Vortices may also form in the corotation region, and can dramatically modify migration behaviour through direct gravitational interaction with the planet. These features become apparent at high resolution, and for all but the highest viscosities there exist significant difficulties in obtaining numerically converged results. The migration of partial gap-opening planets, however, clearly becomes chaotic for sufficiently low viscosities.
At moderate viscosity, a smooth disc-feedback regime is found in which migration can slow substantially, and the migration time-scale observed corresponds to migration being driven by diffusive relaxation of the gap edges.
At high viscosity classical Type~I migration is recovered.
For Jupiter-analogue planets in inviscid discs, a wide, deep gap is formed. Transient Type~II migration occurs over radial length-scales corresponding to the gap width, beyond which migration can stall. Finally, we examine the particle trapping driven by structures left in inviscid discs by a migrating planet, and find that particle traps in the form of multiple rings and vortices can persist long after the planet has passed. In this case, the observation of particle traps by submillimetre interferometers such as ALMA cannot be used to infer the current presence of an adjacent planet.
\end{abstract}




\section{Introduction}

Although planet migration due to gravitational interaction between a planet and a protoplanetary disc is
recognized as one of the possible important processes shaping the architectures of planetary systems,
the best understood models are in the context of protoplanetary discs with accretion flows driven by turbulent viscosity \citep[see][for recent reviews]{2012ARA&A..50..211K,2014prpl.conf..667B}. Recent studies, however, challenge the notion that discs maintain significant levels of turbulence. Observational attempts to detect turbulent broadening of molecular line emission from the outer regions of protoplanetary discs have so far only placed upper limits on the amplitudes of the turbulent motions present \citep[e.g.][]{2018ApJ...856..117F}. Analysis of the vertical mixing of dust in the HL Tau system has constrained the turbulent viscosity parameter in that system to be $\alphaSS \lesssim 10^{-4}$ \citep{2016ApJ...816...25P}. And the very latest theoretical models of the planet-forming regions  of protoplanetary discs ($1 \au \lesssim r \lesssim 10 \au$), incorporating non-ideal magnetohydrodynamic (MHD) effects, indicate that discs remain laminar and accretion is driven by the launching of magneto-centrifugal winds from disc surface layers, and possibly because of the wind-up of large-scale horizontal magnetic fields near their mid-planes \citep{2017ApJ...845...75B,2017A&A...600A..75B}. Hence, it is important to examine the migration of planets in inviscid laminar discs and those with low levels of turbulent viscosity.

In studies dealing with low-mass planets embedded in low-viscosity disc models, an important concept advanced by 
\citet{1984Icar...60...29H}, \citet{1989ApJ...347..490W}, and \citet{1997Icar..126..261W}
was that of the `inertial limit'.
The proposed effect was that when a critical planet mass was reached in an inviscid disc, 
the planet would be able to open a gap, causing migration to stall. These studies assumed that the spiral wakes excited by the planet dissipate at their launch locations, leading to the predicted critical planet mass being small.
A significant refinement of this theory, tackling the details of the location of the planet wake's dissipation
in the disc through its steepening into weak shocks was undertaken by 
\citet{2001ApJ...552..793G}, \citet{2002ApJ...569..997R}, and \citet{2002ApJ...572..566R} \citep[see also the recent work by][]{2018MNRAS.479.1986G}.
These works introduced the term disc feedback, as the general ability of the planet to alter the disc surface density where the planetary wakes dissipate and deposit their energy and angular momentum, with the modified surface density causing first slowing, and eventually stopping, of the planet's migration once it reaches the critical mass.
Numerical experiments by \citet{2009ApJ...690L..52L} and \citet{2010ApJ...712..198Y} (see also the more recent work by \citet{2017ApJ...839..100F} and \citet{2018ApJ...859..126F})
claimed to verify the predictions of \citet{2002ApJ...572..566R}, in that they reproduced the critical mass where planet migration would be halted.
Significantly, these simulations for the first time showed the feedback-driven modification of the disc surface density profile, 
with the planet's migration stalling in a partial gap with a strongly asymmetrical density profile interior and exterior to the planet's orbital location.
Additionally, these works found that in an inviscid or sufficiently low-viscosity disc, the gap edge pressure maxima would be unstable to the Rossby wave instability (RWI), showing a strong tendency to form vortices that exerted significant torques on the planet.
At moderate viscosities, these vortices were suppressed, leading to a smooth feedback-modified migration. At low viscosities these vortices led to rapid and erratic inward and outward migration episodes, as has been examined in detail by  \citet{2010MNRAS.405.1473L}.

The RWI in Keplerian discs can drive vortex formation where a sufficiently large local extremum in 
the radial profile of a quantity $\mathcal{L}\equiv (\Sigma \Omega/\kappa^2)S^{2/\gamma}$, with $\Sigma$ the disc surface density, $\Omega$ the orbital angular velocity, $\kappa$ the epicyclic frequency, $S$ the entropy, and $\gamma$ the adiabatic index,  exists, 
which can be grossly considered as an entropy-modified vortensity, combined with a sufficiently large perturbation and gradient of the quantity $\mathcal{L}$  \citep{1999ApJ...513..805L}.
In the context of locally isothermal discs this quantity reduces to $\mathcal{L} = \Sigma/(2 (\nabla \times \V )_z)$, or half the vortensity.
\citet{2016ApJ...823...84O,2018ApJ...864...70O} have proposed a necessary and sufficient condition for the onset of the RWI.
It is through producing these local extrema in the vortensity profile of the disc where the wake is dissipated that the disc feedback process leads to Rossby wave vortex formation.
The impact of these vortices on planet migration will be a central topic of this paper.

This work aims to extend our understanding of planet migration behaviour in 
low-turbulence wind-driven discs to higher masses than we considered in \citet{2017MNRAS.472.1565M} and \citet{2018MNRAS.477.4596M}, where we focussed on low-mass planets that do not form gaps.
This requires tackling the behaviours introduced by the planet's ability to significantly modify the disc surface density, namely the aforementioned feedback on migration, the formation of vortices and their influence on planet migration. Given the important role vortices play in the migration of planets in inviscid discs, a key element of the work we present here is a resolution study aimed at understanding migration behaviour in the context of numerical convergence. An important conclusion of our work is that such convergence cannot easily be obtained in inviscid two-dimensional simulations of the type we present here, even when simulations approach a resolution of 100 grid cells per scale height.

The contents of the rest of the paper are as follows. We discuss the current state of migration theory in Section~\ref{sec:migration}.
We discuss the relevant physical scales of the problem and the models we examine in Section~\ref{sec:scales},
and in Section~\ref{sec:ConnectingModels} we make a connection between our simple disc models and recent simulations of magnetized discs that incorporate non-ideal MHD effects.
Our numerical methodology is summarised in Section~\ref{sec:methods}.
The results of a range of numerical experiments are reported in Section~\ref{sec:results}, these being 
 feedback modified migration in inviscid discs in Section~\ref{sec:inviscid}, 
 feedback modified migration in viscous discs in Section~\ref{sec:lowvisc}, 
 feedback modified migration in inviscid discs with a vortensity gradient in Section~\ref{sec:vortensitygradient},
 feedback modified migration in advective discs in Section~\ref{sec:advectivedisc}, 
 Type~II migration of giant planets in inviscid discs in Section~\ref{sec:typeII},
a brief examination of  dust drift barriers in inviscid discs in Section~\ref{sec:pebbletrap} and
the formation of particle trapping structures by planets in the outer disc in Section~\ref{sec:yarfm}.
We present a discussion of issues raised by the results in Section~\ref{sec:discussion}, 
and draw conclusions in Section~\ref{sec:conclusions}.

\begin{figure}
\begin{center}
\includegraphics[width=0.9\columnwidth, angle=0]{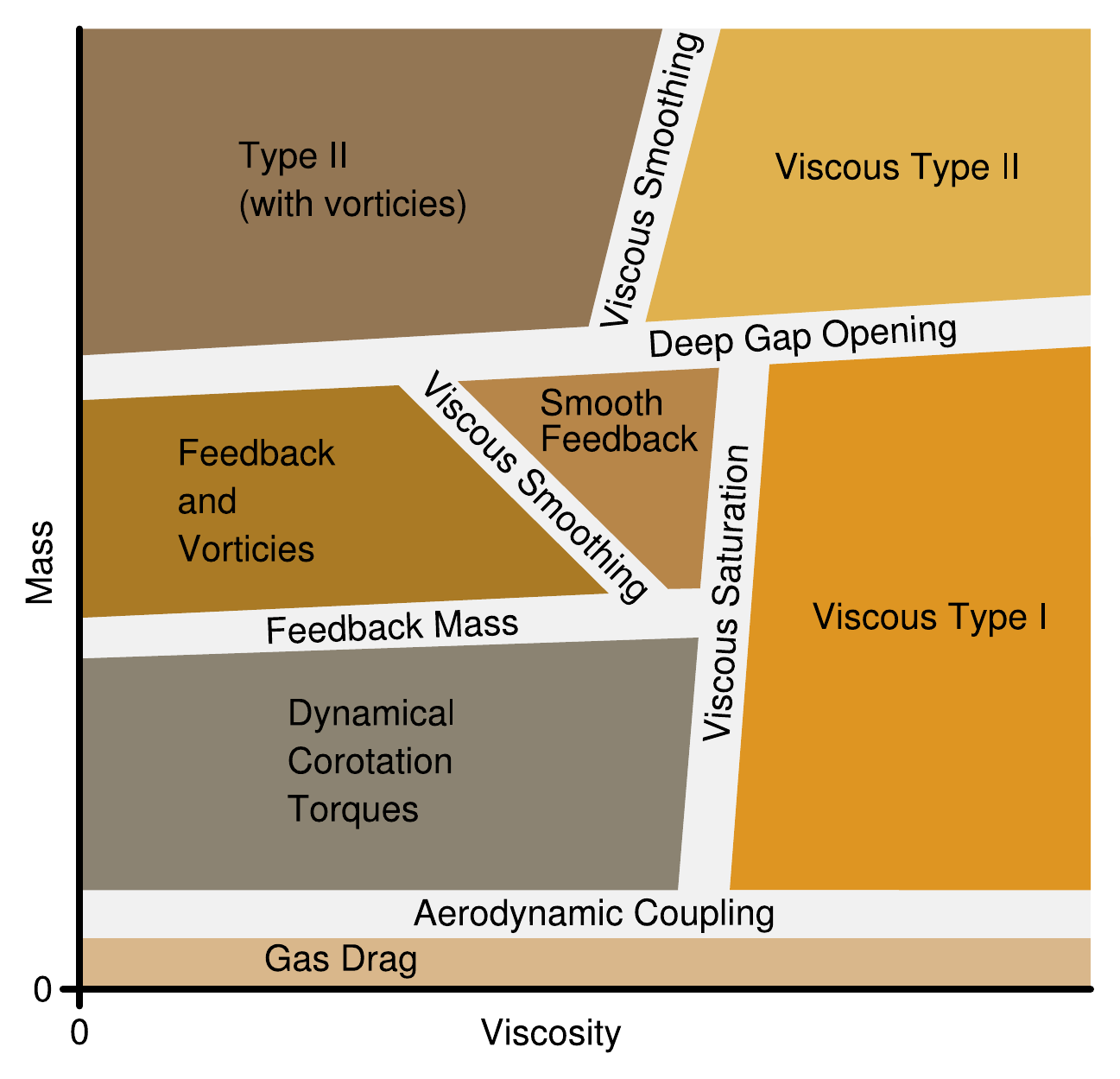}
\end{center}
\caption{Conceptual `city map' of disc migration driving mechanisms in isothermal discs. 
Not all regions are well characterized, and this map does not include thermal effects (and other physics).
Super-Earths growing from planetary cores in the inner dead-zone parts of protoplanetary discs would 
evolve up the left-hand side of the diagram, with final masses around or below the deep gap opening transition.}
\label{fig:migration_org_chart}
\end{figure}

\section{Migration behaviour regimes}
\label{sec:migration}
The landscape of migration behaviour operating under varying physical conditions is complicated.
To motivate our work and place it in context, we review here a subspace of the physical parameters where the collected work in the literature has led to a reasonably complete picture.
In the results of the experiments in this work we will find examples of a number of these behaviours.
Considering only two-dimensional dynamics in globally isothermal power-law disc models of fixed mass, and varying the planet mass and disc viscosity parameters, we suggest as a rough organization the schematic diagram shown in Figure~\ref{fig:migration_org_chart}, with regions of dominant migration mechanisms and transitions between them. Most of the transitions are not sharply defined, and we can best recognize some regimes where the behaviour is clearly of one qualitative type. Moving through Figure~\ref{fig:migration_org_chart} from bottom right through to bottom left in an anticlockwise direction, the various migration regimes can be described as follows.
\begin{description}
\item[\bf Gas drag:] 
The pressure gradient in the disc partially supports the gas against the star's gravity, so in equilibrium 
it orbits at different rates to the Keplerian velocities of orbiting solid particles. The orbital velocity of the gas is usually slower than Keplerian, and embedded solid particles feel a headwind and drift inwards relative to the gas \citep{1972fpp..conf..211W,1977MNRAS.180...57W}. More generally, the drift is directed radially towards higher gas pressures, either radially inwards or outwards, depending on local pressure gradients. Gas drag induced migration can be extremely rapid: metre-sized bodies embedded in a minimum mass nebula at 1 au can reach the star in $\sim 100$ yr \citep{1977MNRAS.180...57W}. For larger bodies, e.g. 10 km-sized planetesimals, drag-induced migration becomes extremely slow and drift times exceed disc life times.

\item[\bf Viscous Type I:] For disc models similar to the minimum mass nebula, gravitational interaction between a planet and the disc becomes relevant for Mars-mass planets and above. Classical Type I migration is driven by the excitation of spiral wakes at Lindblad resonances, giving rise to the Lindblad torque \citep{1980ApJ...241..425G}, and by the interaction between the planet and gas that executes horseshoe orbits in the corotation region, giving rise to the corotation torque \citep{1991LPI....22.1463W}. 
 For conventional disc models, the Lindblad torque is almost always negative, driving inward migration. The corotation torque is normally positive, and can counterbalance the Lindblad torque provided sufficient viscous and thermal diffusion occurs across the horseshoe region, maintaining the entropy and vortensity gradients from which the corotation torque arises \citep{2011MNRAS.410..293P,2014MNRAS.444.2031P}. This corotation torque saturates and switches off in an inviscid disc.

\item[\bf Viscous Type II:] Classical Type II migration applies to planets that open deep gaps in viscous discs. Originally, the planet's migration was envisaged to be locked to the viscous accretion flow of the disc \citep{1986ApJ...309..846L}. More recent studies have demonstrated that the planet and disc flow are not locked, due to the inability of the planet to completely deplete the gap of gas \citep{2014ApJ...792L..10D,2015A&A...574A..52D}, leading to the suggestion that migration occurs at the rate determined simply by the integrated Lindblad torque across the gap region. \citet{2018ApJ...864...77I} have recently suggested that the Type II migration rate can be calculated from the Type I migration rate determined by the surface density at the base of the gap, without specific reference to viscous diffusion rates, based on numerical studies by \citet{2018ApJ...861..140K}. In contrast, \citet{2018A&A...617A..98R}, have shown that Type II migration rates depend explicitly on viscosity (as must be the case if, in the absence of flow from the inner to the outer disc, the location of the gap edge follows the planet as it migrates), and suggest that Type II migration rates are in fact determined by the local viscous smoothing of gap edges, even if a global viscously-driven accretion flow is not present in the disc.

Considerable analytical and simulation effort has been invested in characterising the gap opening mass 
which marks the transition between Type I and Type II migration.
At high viscosities the definition is somewhat more clear than at low viscosities, where partial gap opening over longer time-scales gives rise to what we have denoted as the feedback regime.
The transition from viscous Type~I to viscous Type~II migration has recently been surveyed extensively by \citet{2018ApJ...861..140K}.
The dimensionless parameter 
\begin{align}
K\equiv \left(\frac{\Mp}{M_{\star}}\right)^2 h^{-5} \alphaSS^{-1}
\label{eqn:kanagawa-gap}
\end{align}
has been proposed to control the transition between these migration types by gap opening, with the transition 
occurring smoothly around values of order $K \sim 10$. Here, $\alphaSS = \nu/(c_s H)$, $h$ is the disc aspect ratio, $H$ is the disc scale height, $c_s$ is the sound speed, $\Mp$ is the planet mass and $M_{\star}$ is the stellar mass, and $\nu$ is the kinematic viscosity.

\item[\bf Type II with vortices:] At sufficiently low viscosities, gap edge vortices are formed \citep{2006MNRAS.370..529D,2007A&A...471.1043D}. These induce dissipation
and smear the gap, and results are suggestive of a scenario in which the gap edge settles at a profile with marginal stability to the RWI \citep{2014ApJ...788L..41F, 2017MNRAS.469.3813H,2017MNRAS.466.3533H}. Although the disc surface density begins to be modified by the planet around the feedback mass (defined below), as the gap becomes very wide and clean behaviour more clearly recognizable as Type II emerges. This is a situation that we consider in Section~\ref{sec:typeII}. Our results show that vortex-smoothing of gap edges in an inviscid disc without any accretion flow leads to a transient period of Type II migration over a length-scale comparable to the local disc scale height, which can be sustained with mass flow
from the inner to outer disc.
We do not consider the possible influence of an imposed accretion flow on this type of migration in our suite of runs, but the possible ways that this may change the results are discussed in Section~\ref{sec:scenario2}.

As inviscid conditions ($\alphaSS\rightarrow0$) are approached, the gap opening condition given by equation~(\ref{eqn:kanagawa-gap}) breaks down. The mass scale where the wake begins to shock at the launching point is a useful scale on which to label the gap opening mass in the inviscid limit
\begin{align}
M_1 = \frac{2 c_s^3}{3 \Omega G}\ ,\label{eq:M1}
\end{align}
as introduced by \citet{2001ApJ...552..793G}, where $G$ is the gravitational constant. 
The parameter $M_1$ has the same scaling as the criterion
that the Hill radius of the planet is equal to the disc pressure scale height (sometimes called the ``thermal mass''). 
This is not a very sharp criterion in practice, and the Type II migration arising in very clean gaps, 
as typically considered for a Jupiter-like case corresponds to a planet of mass $\sim 10\, M_1$.
The reader should also note that the various parameters used in the literature for this non-linearity or thermal criterion often differ by small factors of order unity, 
and some with the same scaling are derived through different physical arguments \citep{1993prpl.conf..749L,1996ApJS..105..181K,1997Icar..126..261W,2018SSRv..214...38P}\footnote{The reader may already be familiar with arguments that the gap opening mass ought to correspond to the mass where the Hill radius or Roche lobe radius is equal to the disc pressure scale height.}.

\item[\bf Feedback and vortices:] Where the planet mass is large enough for the wake dissipation to 
alter the local disc surface density profile, and hence alter the Lindblad torques, the feedback slows planet migration. In low-viscosity discs, it has previously been found that vortices can be driven from the altered surface density and rotation profiles, which in turn exert strong and erratic torques on the planet \citep{2005ApJ...624.1003L,2009ApJ...690L..52L,2010ApJ...712..198Y}. Our results indicate that even when this strong, stochastic, vortex-induced migration is absent, weaker vortices still form at gap edges, and their smoothing effect modifies the Lindblad torque such that slow inward migration is sustained. 

A scale for the feedback mass, the mass where the planet's wake dissipation is radially close enough to the planet's location 
to begin to carve a partial gap and feedback on the migration torques, has been derived from linear theory by \citet{2002ApJ...572..566R} as
\begin{align}
M_{\rm F} \simeq 2.5 \frac{\cs}{\Omega G}\left(\frac{Q}{h}\right)^{-5/13} = 3.8 \left(\frac{Q}{h}\right)^{-5/13} M_1 \ , \label{eqn:mfdef}
\end{align}
where $M_1$ is defined by equation~(\ref{eqn:kanagawa-gap}), $\cs$ is the sound speed,  $Q$ is the Toomre $Q$ parameter for the unperturbed disc at the planet location
\begin{align}
Q \equiv \frac{\Omega \cs}{\pi G\Sigma}.
\end{align}
For typically considered planet formation conditions, when $Q$ is significantly greater than unity, as it must be to inhibit gravitational instability, 
and $h$ is of the order of $0.05$ or less, $M_{\rm F}$ is usually less than $M_1$.
This mass was denoted the `stopping mass' in \citet{2002ApJ...572..566R}, but we refer to it here as the feedback mass.
This is because we find in this paper that it in general corresponds more to a mass scale where disc feedback begins to become 
important than  to a mass where migration fully stops in inviscid discs.
This transition has also been termed the inertial limit \citep{1984Icar...60...29H,1989ApJ...347..490W,1997Icar..126..261W} 
under the conception that it represents a mass where migration is halted.

\item[\bf Smooth feedback:] Numerical studies have suggested a viscosity regime exists where feedback can slow or even stop planet migration, where the disc viscosity is sufficient to smooth the disc and prevent vortex formation, but not too large to close the gap \citep{2009ApJ...690L..52L,2010ApJ...712..198Y}. Our results corroborate these earlier studies.

\item[\bf Dynamical corotation torques:] The symmetry of the horseshoe orbits in the corotation region is broken for a migrating planet, and at low viscosities this librating region becomes decoupled from the surrounding disc. As the planet moves, a contrast develops between the vortensity of the gas trapped in the librating region and that in the surrounding disc. The combination of horseshoe asymmetry and vortensity contrast gives rise to a dynamical corotation torque, as described in 
\citet{2014MNRAS.444.2031P}. The influence of the dynamical corotation torque depends on whether or not a radial gas flow exists in the disc, as may be generated for example by large-scale magnetic fields in a dead zone \citep{2017MNRAS.472.1565M,2018MNRAS.477.4596M}.
This effect may in principal continue as long as there is significant mass within the corotation region, even if a gap is partially established.
\end{description}

The conceptual map in Figure~\ref{fig:migration_org_chart} glosses over the role of varying  the disc surface density at a fixed disc temperature (or, equivalently, scale height). As can be seen from equation~(\ref{eqn:mfdef}), the feedback mass in particular is sensitive to the disc mass. If we imagine sitting in a frame moving with a migrating planet, then it is clear that the ability of a planet to alter the local surface density depends on the time-scale required for wake dissipation to open a gap versus the time-scale for the planet to migrate across the gap width (or equivalently, the time for gas to flow into the gap region when the problem is viewed from within the planet frame), with the latter being determined by the disc mass.

In addition to the  migration regimes described above and in Figure~\ref{fig:migration_org_chart}, the 
phenomena of runaway, or Type~III migration may occur when a partial gap (i.e. shallower than a `deep gap') 
can be opened.
In a sufficiently massive disc, a planet may be able to initially migrate fast enough to 
acquire a coorbital mass deficit able to accelerate the planet's 
motion \citep{2003ApJ...588..494M,2004pfte.confE...2A,2007prpl.conf..655P}. 
This form of dynamical torque leads to a transient behaviour, and may occur for planets roughly 
between the feedback mass and deep gap opening masses, for sufficiently massive discs.

\section{Scales}
\label{sec:scales}
To examine the planet--disc interaction problem for planets of the interesting and somewhat narrow 
range where disc feedback plays a role, planets with the correct mass must be chosen.
In this paper, we denote the disc pressure scale height by $H$ as a dimensional quantity, and use
the notation $h\equiv H/\rp $ for the non-dimensional disc aspect ratio at the orbital radius of the planet $\rp$.
We adopt a canonical disc surface density model by defining 
a minimum mass solar nebula (MMSN) surface density profile:
\begin{align}
\Sigma &= \Sigma_0 \left( \frac{r}{r_0} \right) ^{-\alpha}\ , \\
\Sigma_0  &= 1700 \ \mathrm{ g\ cm^{-2}} = 1.9\times 10^{-4}  \ \mathrm{ M_\odot\ au^{-2}} \ ,
\end{align}
with $r_0=1\ \mathrm{au}$ and $\alpha=3/2$.
To set the sound speed, $c_s$, we define the aspect ratio to be $h=0.035$ at $1\ \mathrm{au}$, 
which implies $c_s=1.05\ \mathrm{km\ s^{-1}}$, which we adopt globally.
Thus, the aspect ratio of the disc varies radially, and the sound speed is globally constant.
These choices were made for two reasons. First, the main focus of this paper is the role of disc feedback in changing planet migration, and employing $\alpha=3/2$ gives uniform vortensity in the disc, and hence no corotation torque. Secondly, in order to maintain a set up as simple as possible, with the fewest number of free parameters, a constant sound speed gives a barotropic equation of state, preventing baroclinic effects from influencing the already complicated results  (e.g. vortices may be sourced by baroclinic effects, in addition to the RWI). We note here that models are presented with $\alpha =1/2$ later in the paper, to examine how a vortensity gradient influences the results, but the other parameters defined above are kept constant.

The relevant scale for the planet mass which determines when disc feedback plays a role is the feedback mass, as discussed in the previous section.
To determine this mass, \citet{2002ApJ...572..566R} analysed analytically the effects of shocking and damping of density waves launched by a planet.
After finding a stationary solution for the disc surface density in the frame of a migrating planet, 
the planet mass for which this solution breaks down, such that no stationary solution for a partial gap exists, was interpreted as a condition where the planet will open a gap, and stop migrating.
The critical mass for stopping migration $M_{\rm F}$ derived from the theory of  \citet{2002ApJ...572..566R}  is 
as given by equation~(\ref{eqn:mfdef}), or equivalently
\begin{align}
M_{\rm F} \simeq 3.8 \left(Q \frac{\rp}{H}\right)^{-5/13} M_1\ ,
\end{align}
in terms of $M_1$ (equation~\ref{eq:M1}).
In a disc with twice the surface density of the MMSN-like disc defined above, which we take as our fiducial model in this work, so that
$\Sigma = 2 \times (1.9\times 10^{-4} ) (r/1\ \mathrm{au})^{-3/2}\ \mathrm{ M_\odot\ au^{-2}} $
at $1\ \mathrm{au}$, and with aspect ratio $h=0.035$, the feedback mass is  
$M_{\rm F}=8.0\times10^{-6} \ \mathrm{M_\odot} $ or $2.7\ \mathrm{M_\oplus}$. 
In the same disc and planet position the $M_1$ scale for the wake to shock on launching is $2.86\times10^{-5} \ \mathrm{M_\odot} $ or $9.5\ \mathrm{M_\oplus}$.
It is important to note that the \citet{2002ApJ...572..566R} analytical theory does not directly predict that the planet must stop migrating. 
The prediction of migration stopping is inferred from the breakdown of the density wave dissipation solution for a moving planet, 
that is when steady-state solutions for a moving planet cease to exist, the interpretation assigned to the scenario 
is that the planet is able to open a gap in the (inviscid) disc, transition to Type~II migration, and hence stop.
No account was made in the one-dimensional treatment for the possibility of 
two-dimensional instabilities occurring as the planet modifies the disc surface density and vortensity profile.
Indeed, \citet{2009ApJ...690L..52L} predicted, and \citet{2010ApJ...712..198Y} found 
Rossby wave driven vortex formation in this situation.

At large viscosities, the memory of the flow is short, as viscosity can quickly diffuse away any perturbation 
caused by the planet to the vortensity or surface density. As viscosity decreases, the time-scale for the 
disc to relax back to the initial equilibrium becomes longer and longer.
The diffusive time-scale then, for features on a length $\ell$ due to a viscosity $\nu$, is $\ell^2/\nu$.
The disc has an accretion time-scale, but as this accretion flow likely only involves the upper layers of the disc in a wind-driven disc scenario
this time-scale may not be strongly relevant to the midplane dynamics.

The disc has a lifetime  time-scale which is by definition relevant to disc--planet interactions, and should be of the order of several Myr.
Additionally the planet itself has a growth time-scales as it accretes solids and gas in what are possibly several separate stages.

The modelling approach most commonly considered for disc--planet interactions assumes that the flow time-scales are short 
compared to the planet growth and disc lifetime time-scales.
For very low-viscosity discs, this assertion is not as trivially true as it is for viscous accretion discs.
Conventionally, the technique employed for studying planet--disc interactions 
has been to introduce a planet potential either instantaneously
 or over a short time-scale (a handful of orbits) into an unperturbed disc.
For a viscous disc, it would also be possible to start with a viscous  steady-state for a fixed or moving planet potential.
Such an approach would be reasonable for testing if a given steady-state is stable, but it is more interesting in a planet formation 
 context to find states that systems will be inexorably driven into as they evolve.
 Moreover, if the planet motion is not smooth and varies on short time-scales or oscillates, such simply defined steady states are not possible.
 Thus, a reasonable experiment is still to introduce a planet potential over a short time-scale and examine the long-term evolution of the planet motion
 as a guide for how forming planets move in protoplanetary discs.

\section{Connecting simple models to magnetized discs}
\label{sec:ConnectingModels}
Simulations of magnetized discs incorporating non-ideal MHD show two distinct behaviours, depending on whether the vertical component of the magnetic field is aligned or misaligned with the disc's rotation vector \citep[e.g.][]{2017ApJ...845...75B,2017A&A...600A..75B}, where the Hall effect induces the aligned/misaligned dichotomy \citep{1999MNRAS.307..849W,2001ApJ...552..235B}. In anti-aligned scenarios, a magneto-centrifugal wind is launched from disc surface layers (with the assistance of heating of the wind), and essentially all of the back reaction onto the disc occurs in an ionized surface layer with a small actively accreting column density  $\Sigma_{\rm Active} \sim 1$ g cm$^{-2}$. Gas at lower latitudes is rendered magnetically dead by Ohmic resistivity and ambipolar diffusion, and remains laminar without accreting. The base of the active layer sits at height $z > 4 H$, whereas the corotation region of a low-mass planet embedded in a disc extends up to approximately $z = 3 H$ \citep{2015ApJ...811..101F}. Hence, in this work we assume that a 2D inviscid disc model acts as a good proxy for a 3D disc with accretion occurring at the surface only for planets that remain largely embedded. The formation of a significant gap may change this, but the very small column density associated with the actively accreting layer will mean that it only weakly interacts with the planet as it passes into the gap region.

In aligned scenarios, the Hall shear instability \citep{2008MNRAS.385.1494K} generates large-scale horizontal magnetic fields at mid-latitudes, and these can diffuse to the mid-plane and wind up due to Keplerian shear, generating strong mid-plane accretion flows \citep{2014ApJ...791..137B, 2014A&A...566A..56L} in addition to the surface accretion generated by the magneto-centrifugal wind. Here, a large fraction of the column density of the disc is actively accreting. In this work we follow our previous investigations of what happens to planets embedded in discs with aligned magnetic fields, by modelling the disc as a 2D object in the ($r$,$\phi$) plane with a body force mimicking Lorentz forces \citep{2017MNRAS.472.1565M,2018MNRAS.477.4596M}.

In both the aligned and misaligned scenarios, we note that Ohmic resistivity is extremely effective at diffusing any small scale magnetic field perturbations, caused either by embedded planets or by vortices. Hence, treating magnetic effects with a body force that drives a global accretion flow should be a good approximation to the aligned case, and for the misaligned case treating the disc as inviscid and non-accreting should also be a good approximation.

The inclusion of viscosity in the parameter survey is motivated by its smoothing effect on the flow, both as a source of smoothing of planet-induced gaps, and also the smoothing of pressure bumps such that vortex formation via the RWI is quenched. It is not motivated by any particular physical model for turbulent viscosity or by a strong belief that discs are turbulent in nature.

\section{Methods}
\label{sec:methods}
The fiducial model used in this work is a 2D radial-azimuthal $(r,\phi)$ vertically integrated globally isothermal MMSN-like disc, without self-gravity and
with Plummer-sphere smoothing of the planet potential in a modified version of the {\sc FARGO3D} code \citep{2000A&AS..141..165M,2016ApJS..223...11B}. 
This version of {\sc FARGO3D} includes adaptions for queue managed computing clusters, hybrid OpenMP/MPI parallelism, and a particle module.
The grid covers the radial interval from $r=[0.3,2.0]$ with a variable resolution and wave-damping 
zones on the boundaries described in Appendix~\ref{app:grid}.
We include the indirect terms arising from the planet and disc and a correction to the forces applied to the planet due to the lack of disc self-gravity \citep{2008ApJ...678..483B}.
The softening length of the planet potential is $0.4H$.
 For all planets except the giant ones considered in Section~\ref{sec:typeII} the entire simulation domain is included in the calculation of the forces acting on the planet.
In the {\sc FARGO3D} code the planet's trajectory is integrated with a fifth-order Runge--Kutta scheme sub-cycled five times in the gas solution time step, which is small compared to the time-scale of encounters with gas vortices.
The disc is initialized in exact numerical equilibrium with respect to the discretized radial centrifugal, gravitational, and pressure forces.
Unless otherwise noted, in all models the planet mass was ramped up over two orbits before the planet is released and begins to follow the trajectory dictated by the disc torques. 
The calculation is performed in the guiding centre reference frame.

\section{Results}
\label{sec:results}
\subsection{Inviscid MMSN-like discs}
\label{sec:inviscid}
\begin{figure}
\includegraphics[width=\columnwidth]{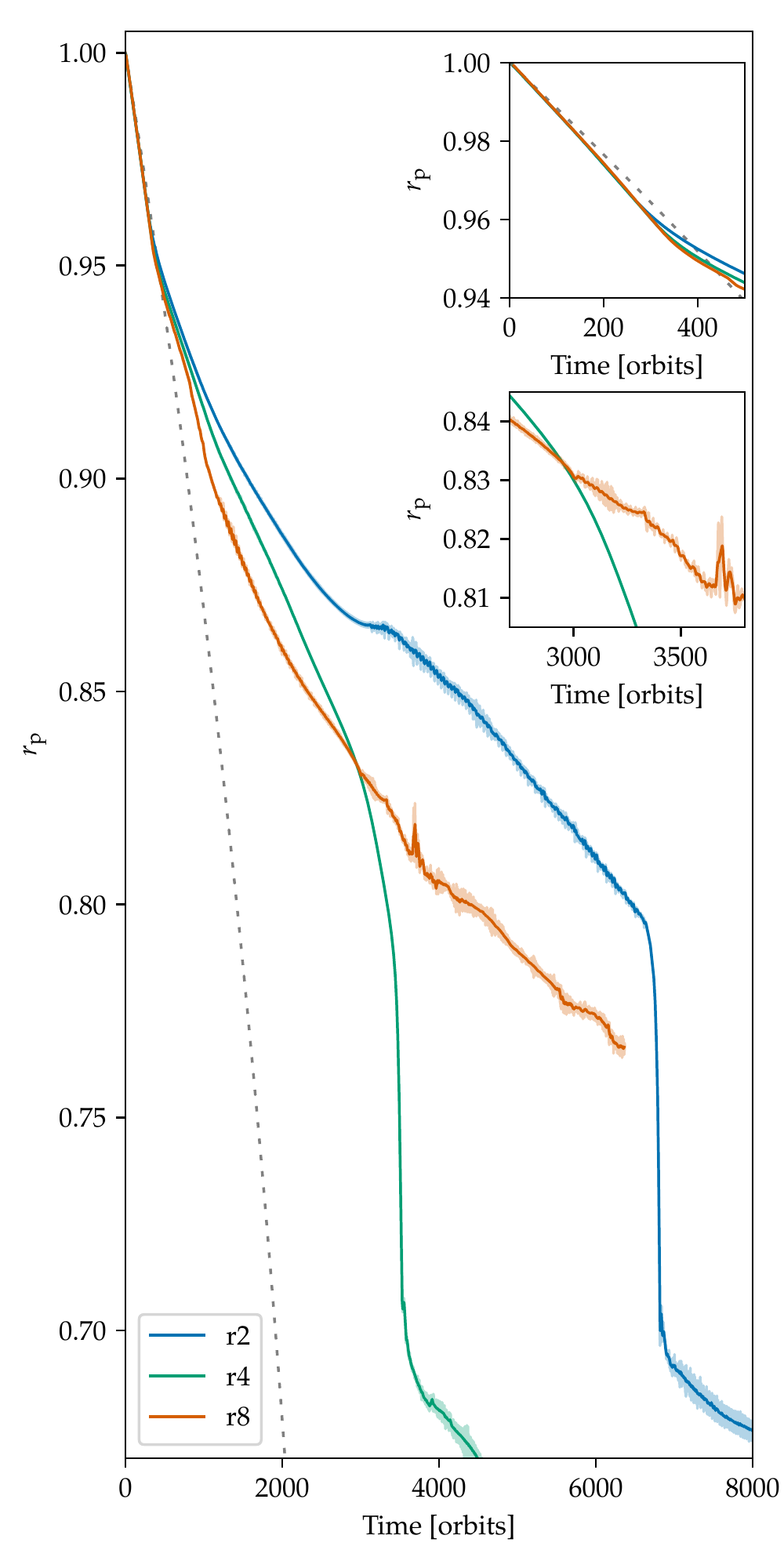}
\caption{Planet with mass $q=1.25\times10^{-5}$ in an inviscid disc, resolution study. 
Light coloured traces show the instantaneous trajectory, 
and dark coloured traces show a 16 orbit running average, adopted as a standard throughout the plots in this paper. The lowest resolution `r2'  (23 zones per scale height) trajectory shows a 
halting of migration just before 400 orbits, after which vigorous vortices lead to a resumption of inward migration.
Grey dashes: Type~I migration rate from \citet{2011MNRAS.410..293P}.
}
\label{fig:gap16}
\end{figure}

\begin{figure}
\includegraphics[width=\columnwidth]{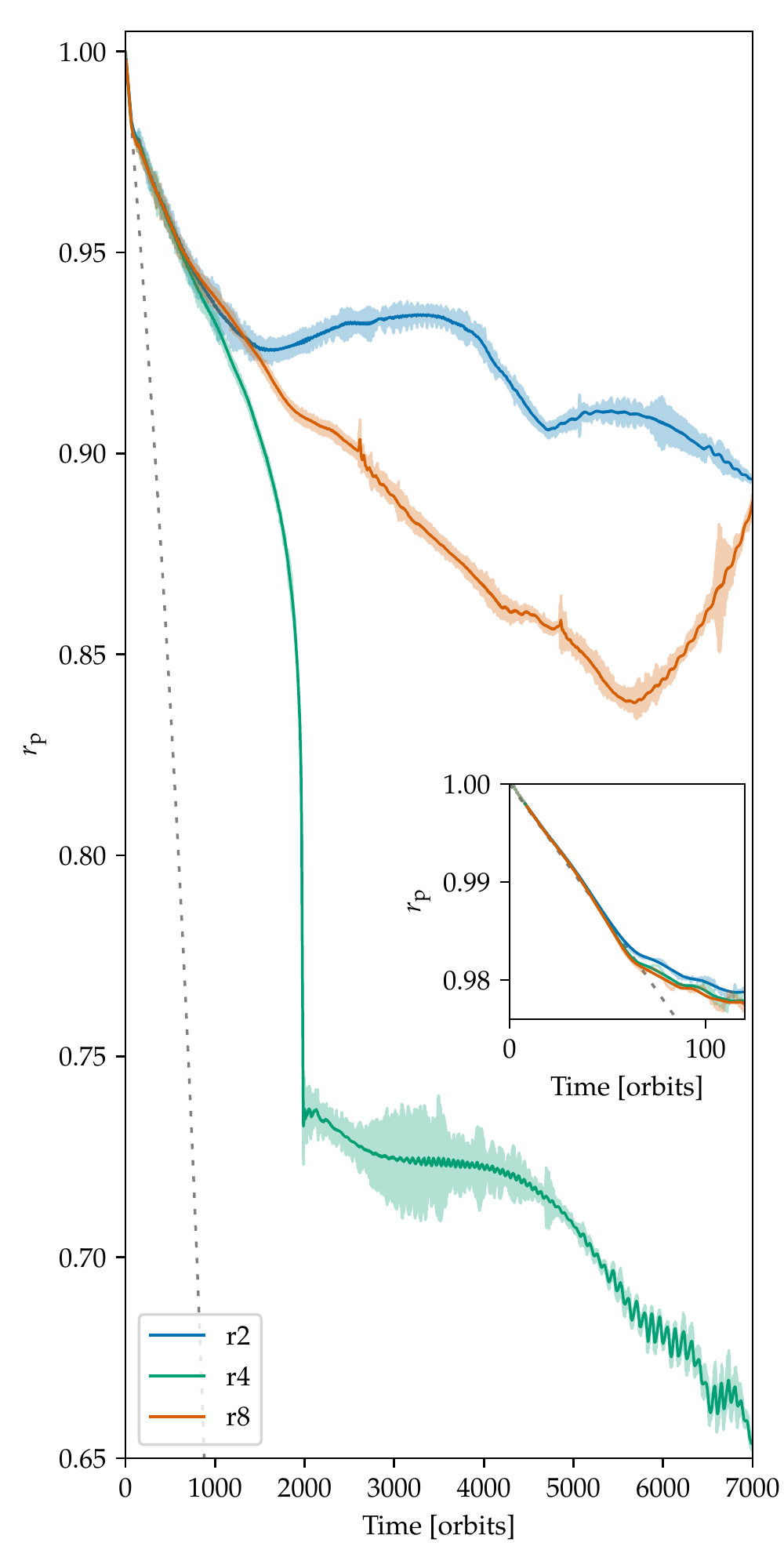}
\caption{$q=3\times10^{-5}$ inviscid disc resolution study.
Light coloured traces show the instantaneous trajectory, 
and dark coloured traces show a 16 orbit running average as standard in this paper. 
The dashed line shows the Type~I migration trajectory.
}
\label{fig:gap29}
\end{figure}

\begin{figure*}
\includegraphics[width=\textwidth]{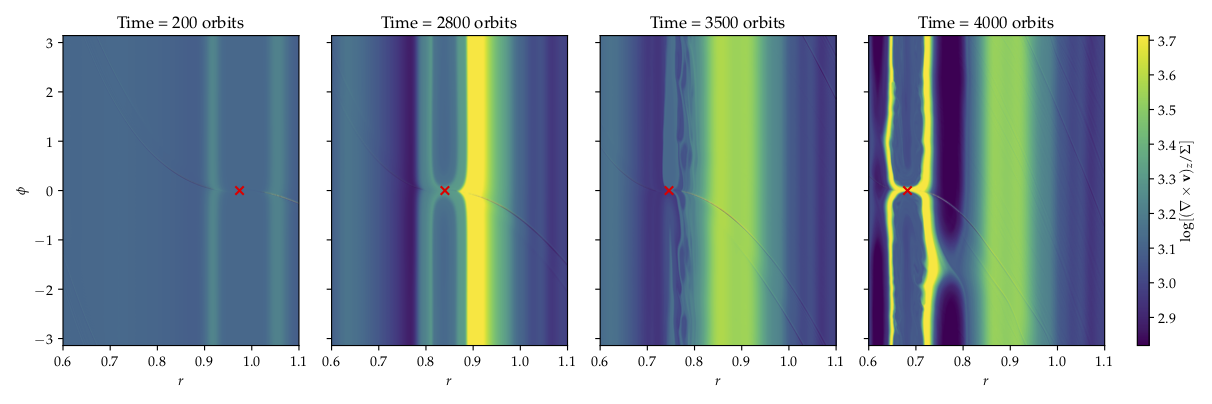}
\caption{
Vortensity evolution before, during and after rapid inward migration episode for $q=1.25\times10^{-5}$ at resolution r4.
The instantaneous planet position is shown by the red $\times$ mark. In the third panel the teardrop-shape libration 
region with high-vortensity material in front 
(above on the page) of the planet characteristic of Type~III migration can be observed, along with a flow through stream of low-vortensity material to the rear
(below on the page). This stream can be seen then spawning RWI driven vortices as it mixes with the material with contrasting vortensity. 
In the third and fourth panels the secondary wake appears within the plots on the right-hand side.}
\label{fig:gap16r4_vort_jump}
\end{figure*}

\begin{figure*}
\includegraphics[width=\textwidth]{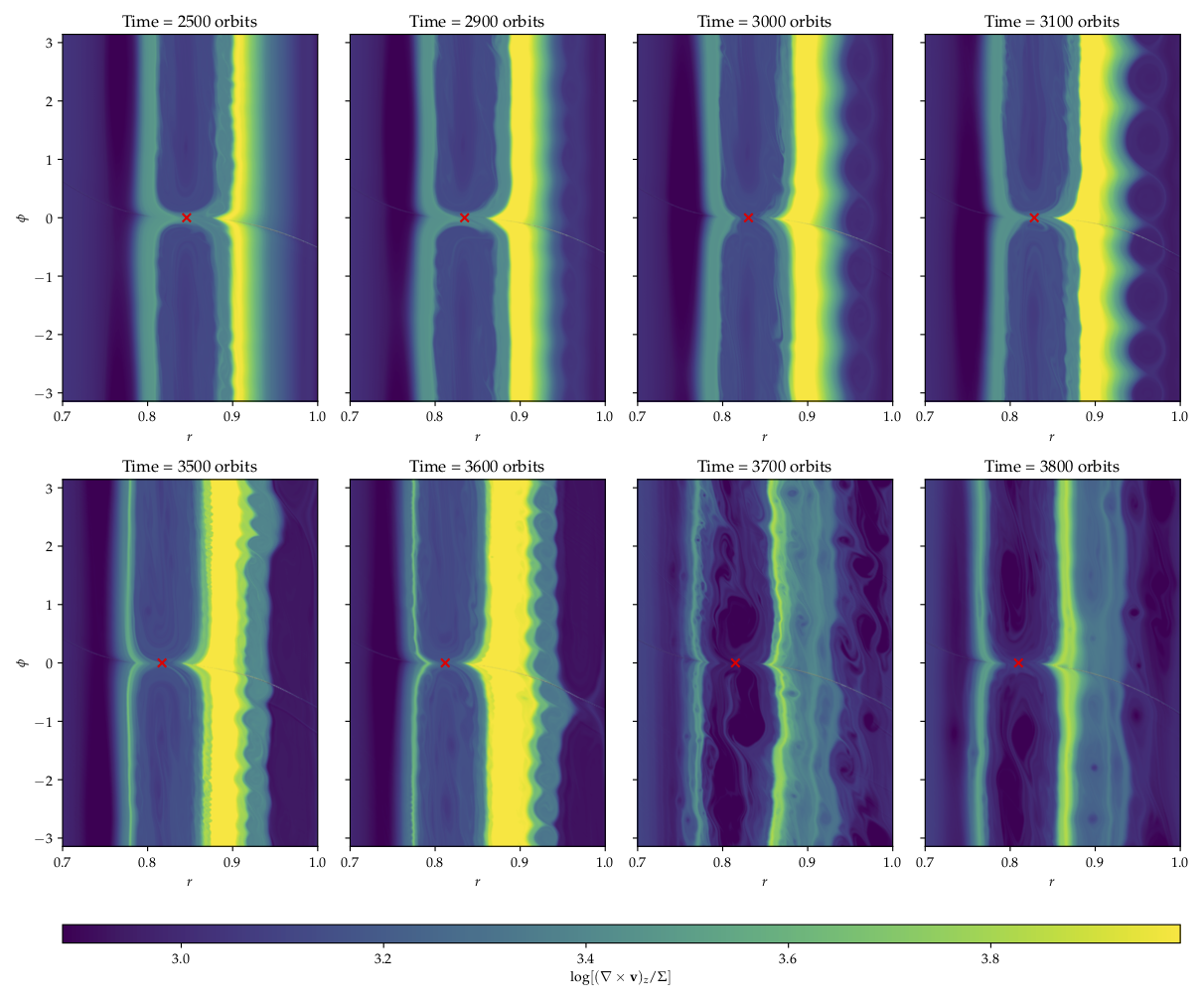}
\caption{Top row: Vortensity evolution during the `failed' transition to fast inward migration for $q=1.25\times10^{-5}$ at resolution r8.
The growth of a set of RWI vortices at the outer gap edge is apparent.
Bottom row: Centred on a burst of RWI activity, resulting in the significant reduction of the peak vortensity.
The instantaneous planet position is marked by the red $\times$.
}
\label{fig:gap16r8_vort_fail}
\end{figure*}

The first set of simulations aims to test the migration behaviour of planets more massive than $M_{\rm F}$ in globally isothermal discs with no vortensity gradients, that is $\Sigma\propto r^{-3/2}$.
Importantly, this particular choice implies that initially the planet experiences no corotation torque, 
although once the surface density profile has been altered, one may arise.
The disc used is the 2 MMSN-like disc detailed in Section~\ref{sec:scales}. 

First, simulations of the migration of a planet with mass $q=1.25\times 10^{-5}$ ($4.2\ \mathrm{M_\oplus}$, corresponding to $1.5\  M_{\rm F}$) were run at three grid resolutions, specifically with grid functions r2, r4, r8 (Appendix~\ref{app:grid}). 
These have variable radial  spacing, approximately  $23$, $47$, and $93$ zones per scale height at the initial planet position of $r=1$.
In total these grids consist of $(n_r,n_\phi) =1243  \times \ 4189$,  $(n_r,n_\phi) = 2484\ \times \  8378$, and $(n_r,n_\phi) = 4966\ \times\ 16755$ cells respectively.
For the planet mass used, the corotation region half-width $\xs \approx 1.2 \rp \sqrt{q/h}$ \citep{2006ApJ...652..730M} is $0.02$, 
and the horseshoe libration time is initially approximately $60$ orbits.
The planet trajectories are shown in Figure~\ref{fig:gap16}.
Once the planets are released after two orbits, the initial evolution for $\sim 300$ orbits agrees well between the simulations at the three resolutions 
and the planet initially migrates at a rate close to that
expected for Type I migration due to Lindblad torques alone \citep{2011MNRAS.410..293P}.
After this period, the disc feedback effect begins to impact on the migration rate, and it slows as the wake is able to open a partial gap and build asymmetrical surface density structure. The gap starts to form inside and outside the planet as a pair of symmetric grooves at the locations where density waves are launched and damp, at a distance $\gtrsim H$ from the planet. Gas in the horseshoe region, however, is also evacuated diffusively because vortices form at the gap edges.

The migration behaviour in Figure~\ref{fig:gap16} qualitatively agrees with the initial evolution of planets in inviscid discs shown in \citet{2010ApJ...712..198Y}.
During this phase, vortices and perturbations to the migration rate are seen for all resolutions, although they grow more rapidly at higher resolutions.
In this phase, the migration rate is initially higher at higher resolution.
The migration is invariably inwards, although details vary depending on the simulation resolution.
At the lowest r2 resolution the planet is observed to slow, 
and migration briefly halts after roughly 3000 orbits. 
However, the eccentricity quickly increases, and the planet resumes inward migration. 
This change is the result of a burst of vortex activity at the partial gap edges.
From roughly 4000 to 7000 orbits the planet migrates inwards, while continuously jostled by vortices which mainly reside at the vortensity peaks to the inside and outside of its orbit. 
This period of jostled but steady migration is terminated by a rapid, inward Type~III migration episode, 
which drives the planet into the high-density ring previously formed by dissipation of the inner wake inside its orbit. Once reaching this point, the fast migration quickly ends, and the planet migrates more slowly while carving a new partial gap.

At twice the resolution, on grid r4, the planet never slows to a near stop. 
Instead, as feedback modified migration sets in at roughly 400 orbits, the vortices that grow in the vortensity peaks are able to keep the planet moving inwards, until it slowly begins to climb into the density peak it has driven in front of itself, and transitions into an episode of rapid inward Type~III migration.
The details of this episode are discussed in Section~\ref{sec:rapidinward}.

Doubling the resolution again, the simulation on grid r8 shows the planet again being affected 
by the presence of vortices as soon as disc feedback sets in. 
Although the migration rate varies, it never achieves a Type~III episode.
Figure~\ref{fig:gap16} includes a zoom-in panel (middle right) detailing a time with two episodes to be discussed in Section~\ref{sec:rapidinward},
where at roughly 3000 orbits the planet appears to start, but fails to make a transition into a Type~III episode, and at roughly 3600 orbits a significant burst of vortices occurs.
Overall, the key aspect of these simulation outcomes is that despite the planet mass being above the feedback mass, and being in an inviscid disc, the planet migration does not halt.

A second set of runs, with a planet mass $q=3\times 10^{-5}$ ($10\ \mathrm{M_\oplus}$) is shown in 
 Figure~\ref{fig:gap29}.
Again, the same disc initial condition and the same three grid resolutions r2, r4, r8  (Appendix~\ref{app:grid}) are used.
This planet is roughly at the thermal mass $M_1$ (which is $q=2.86\times10^{-5}$) and corresponds to $3.5\ M_{\rm F}$.
As $M_1$ scales with orbital frequency it decreases with radius in this isothermal disc,  and at $r=0.85$ is $22\%$ lower
 ($q=2.24\times10^{-5}$).
Hence, the planet moves deeper into the gap-opening regime as it migrates inwards.
At this mass, the planet has a corotation region half-width $\xs \approx 0.035$ and the libration time is initially $\sim 38$ orbits.
This configuration is the closest to that of \citet{2010ApJ...712..198Y} run.
Indeed, the lowest resolution r2 has qualitative similarities, in that the planet is able to drive significant vortices, giving it a variable eccentricity, and after an initial halt wanders slowly both inwards and outwards.
Doubling the resolution to grid r4, the behaviour qualitatively changes. 
Instead of initially undergoing a feedback based halting, the planet enters a feedback enabled Type~III episode, 
terminating once the planet has passed the density peak previously driven in front of it.
The planet proceeds to create a new gap and drive vortices, and then proceeds to resume its inward migration.
At the highest r8 resolution, the results differ qualitatively again. Once feedback and the associated vortices have set in after about 500 orbits, the migration rate is roughly constant for 5000 orbits, while the planet migrates inwards.
During this phase, the migration is briefly interrupted by short bursts outwards, as seen in the lower mass case.
However, at the time 5500 orbits, a new phenomenon emerges and the planet begins to migrate outwards instead of inwards.
This novel behaviour arises because of the formation of a coorbital vortex with which the planet interacts, and will be examined in depth in Section~\ref{sec_corovort}.

The results of these inviscid simulations are varied, but this is a consequence of the attempt to simulate a flow with the lowest possible viscosity, which is inherently resolution-dependent.
At no point does the resolution study suggest that planets above $M_{\rm F}$ in an inviscid, isovortensity disc will halt their inward migration and settle long-term at one position, in contradiction to the simplest interpretation of earlier analytical results. This appears to be primarily due to RWI enabled vortices, a secondary instability of the feedback-modified disc profile which inherently cannot be captured by the one dimensional analytical calculations. Our interpretation is that these vortices introduce dissipation into the flow and smooth the gap profiles, resulting in a net Lindblad torque that continues to drive inward migration. Direct gravitational interaction between a planet and vortices, however, can also significantly modify the orbital evolution in a stochastic manner, and this renders the migration behaviour of partial gap-forming planets in discs with very low-viscosity a fundamentally  chaotic phenomenon.
That is, a small change imposed on the system at one time may result in a large change in the state of the system at a later time. 
 The distribution of possible migration outcomes can then only be understood through a statistical survey based on a large suite of simulations. Consideration of additional phenomena beyond the planet--disc interaction such as planet-planet interactions may then also be required to form a complete picture of how migration influences the final architectures of planetary systems.

We further comment that continuing to increase the resolution to determine if results can be obtained that are at least qualitatively converged, if not quantitatively so, is likely to be a futile exercise that is not physically well motivated. At the highest resolutions the simulations generate structures that are on the scale of $H$, and hence the problem starts to become somewhat ill-posed in two dimensions.

\subsubsection{Rapid inward migration episodes}
\label{sec:rapidinward}
Previous studies of intermediate mass planet migration in low-viscosity discs have 
found intervals of rapid Type~III migration \citep{2009ApJ...690L..52L,2010ApJ...712..198Y}.
Some of the models in this paper also produce these events, although the details of the phenomena differ qualitatively. 
For example, in \citet{2010ApJ...712..198Y} Type~III episodes drove a planet in an inviscid disc both inwards and outwards, 
whereas here only inward episodes were observed.

When Type~III migration episodes occur in these models, they initiate through the planet ingesting high-density 
inner gap-edge material into the downstream (rearward) horseshoe turn.
Figure~\ref{fig:gap16r4_vort_jump} shows the configuration of the fluid vortensity $(\nabla\times\V)_z/\Sigma$ before,
during and after the Type~III episode experienced by the $q=1.25\times10^{-5}$ mass planet at resolution r4 in an inviscid disc.
The planet is slowly and smoothly able to pull low-vortensity material into the rearward horseshoe turn, leading to a strong asymmetry 
with the strongly distorted libration island containing higher vortensity material from the gap. 
The planet proceeds inwards, slowing rapidly after passing the vortensity minimum (a surface density maximum).
During the rapid inward migration phase, the flow-through stream behind the planet emerges as a thin stream of low-vortensity material in a higher vortensity medium, and this sharp contrast leads to the formation of vortices at the edge of the libration island through RWI.
These vortices, evolving in two dimensions, merge into one large vortex seen exterior to the planet in the final panel. After the rapid inward migration phase, the corotation region and libration island have a much more mixed vortensity structure.
 
In simulations in the inviscid limit, the rapid inward migration episode behaviour has a strong dependence on resolution. 
At the highest r8 resolution, neither the $q=1.25\times10^{-5}$ nor $q=3\times 10^{-5}$ mass planet experiences rapid Type~III migration episodes, but it should be noted that we are not able to say definitively whether or not this is because we have not run the r8 simulations long enough to see one, or if instead the Type~III episodes do not occur at higher resolution because new phenomena emerge which prevent them. An example of a scenario which appears to be the failed initiation of a Type~III episode exists in the $q=1.25\times10^{-5}$, r8 inviscid simulation.
The planet trajectory during this period is shown in the middle-right inset panel of Figure~\ref{fig:gap16}, 
where the incident occurs from roughly 2900 to 3000 orbits.
Figure~\ref{fig:gap16r8_vort_fail}, upper row, shows maps of the vortensity distribution before, during and after this time.
In the first two panels (Time = 2500 orbits and Time = 2900 orbits) the planet can be seen moving towards the inside edge of the gap, 
and a wide high-vortensity region is present to the outside.
However, instead of initiating a Type~III episode by interacting with the gap edge, the planet's migration is disrupted by the growth 
of the RWI driven vortices forming at the outer gap edge.
A feature which we commonly observe in the evolution of RWI modes can be seen clearly here --- that the dominant mode number 
decreases as the instability grows, decreasing from $m=9$, to $m=7$, between the second and third panel. 
 This variation of the most unstable radial wavenumber and subsequent vortex merging process in RWI has been studied in detail by \citet{2016ApJ...823...84O,2018ApJ...864...70O}. 
Here, instead of a clear flow-through stream being established as in Figure~\ref{fig:gap16r4_vort_jump}
the oscillations of the planet result in the stream being cut off. 
Later, as the vortensity of the gap grows, growth of RWI on the outside edge of the libration island and the outside edge of the gap leads to a burst 
of vortices which cause outward and inward oscillations of the planet orbital location around 3700 orbits.
The lower row of Figure~\ref{fig:gap16r8_vort_fail} shows this sequence, where first the growth of RWI modes can be seen in the first two panels, 
and the immediate aftermath of the burst of vortices is apparent in the Time = 3700 orbits panel, where the wide strip of high vortensity 
in the gap to the outward side of the planet has been significantly smoothed and relaxed by the action of the instability.
In the final Time = 3800 orbits panel, the process of merging of the vortices and regularization of the flow can be seen.
This relaxation process, where the vortensity peak is rapidly smoothed by the action of RWI vortices has not been observed at lower resolution.

For the higher $q=3\times 10^{-5}$ mass planet, the same transition in qualitative behaviour between the r4 and r8 resolutions 
can be observed in Figure~\ref{fig:gap29}. 
Evidently increasing the grid resolution, and thereby decreasing the numerical diffusive and viscous effects in the simulations, tends to impede Type~III episodes. This seems to be due to the fact that enhanced vortex formation at high resolution increases the jostling of the planet, removing the smooth conditions that precede Type~III episodes at lower resolution.

Despite the qualitative agreement in the trend at highest resolution simulations for the two planet masses, 
this phenomenology is sufficiently qualitatively different from that at lower resolutions (r2, r4) that we have little confidence that the highest resolution 
simulations (r8) are closely approaching their full ultimate limit of behaviour. 
However, that the additional resolution allowing additional smaller scales in the flow with more virulent instability prevents the clean flow structures 
which appear to enable Type~III episodes suggests that considering even further small scales will not allow the return of the 
flow to a smooth state as seen at low resolution.
However, at the r8 resolution, with $93$ zones per scale height, significant vortical flow features much smaller than a scale height appear 
in many places in the flow. 
As this  is the case, the reduction of the problem to two dimensions must be reconsidered in the future, 
as such flow features can behave in significantly different ways when the vertical dimension is introduced.
Specifically, the stability, interaction, or merging of tightly wrapped vortices can be very different in three dimensions at scales below a scale height.
This step will however require a significant investment in computational resources.

\subsubsection{Outward migration of a $q=3\times 10^{-5}$ planet}
\label{sec_corovort}
\begin{figure}
\includegraphics[width=\columnwidth]{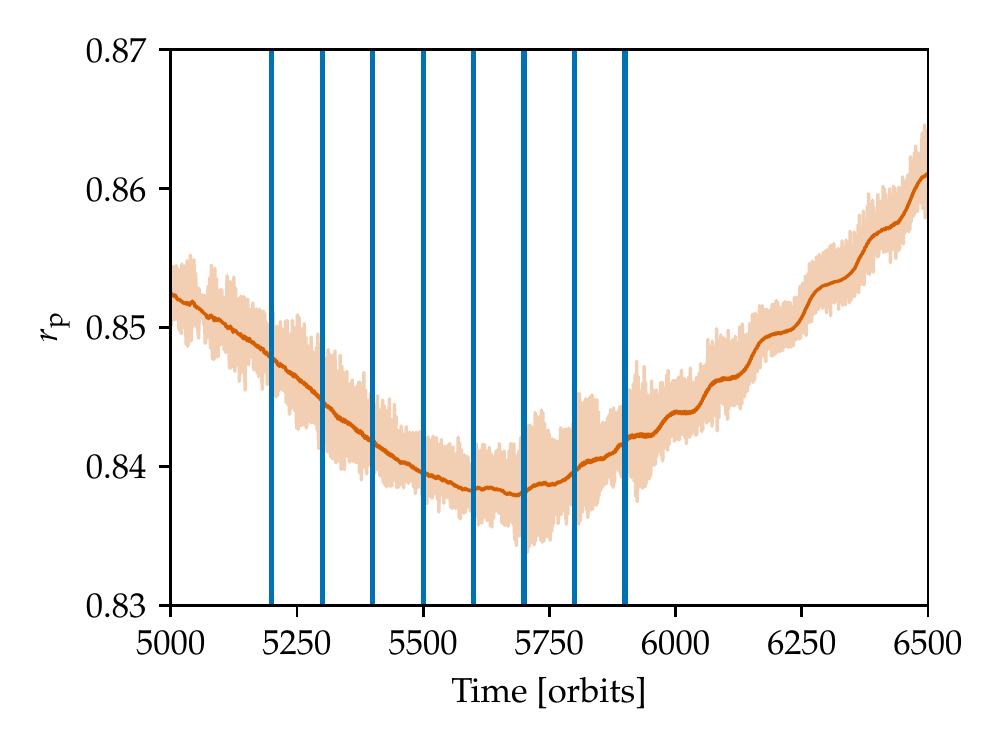}
\caption{Late time transition from inward to outward migration in the $q=3\times10^{-5}$, r8 inviscid disc run.
Vertical lines mark the times of the panels of vortensity and surface density shown in Figure~\ref{fig:gap29_sigma_turn}.}
\label{fig:gap29_turn}
\end{figure}

\begin{figure}
\includegraphics[width=\columnwidth]{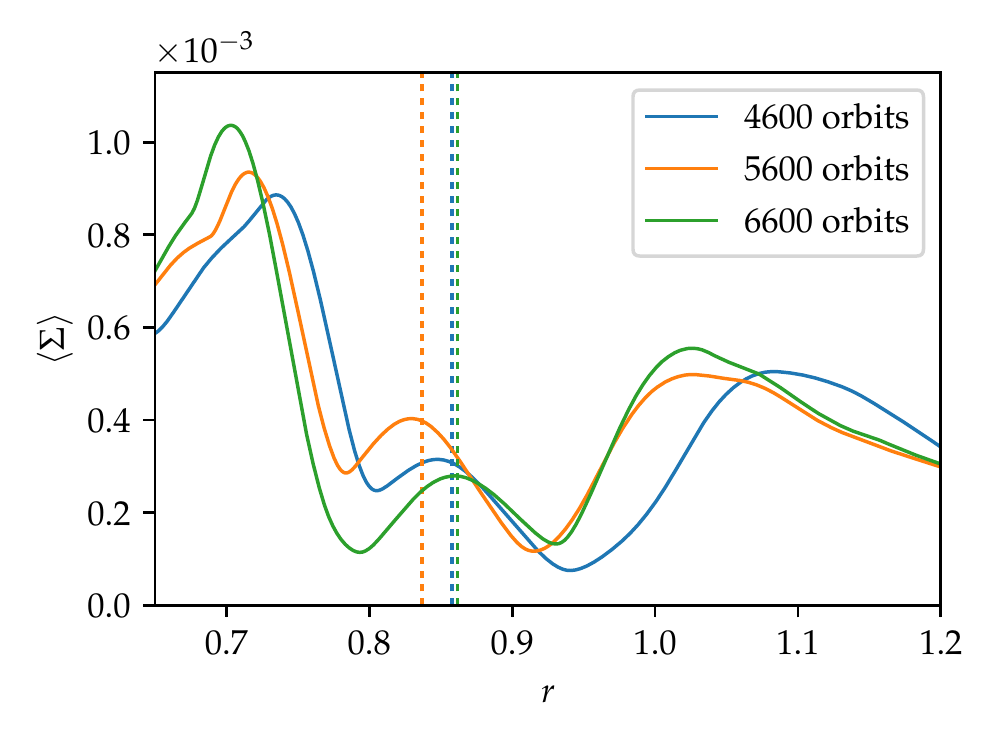}
\caption{Azimuthally averaged surface density in the $q=3\times10^{-5}$, r8 inviscid disc run at 
times bracketing the migration reversal.
The planet position at each time is given by the vertical dashed lines.
}
\label{fig:gap29r8_profiles}
\end{figure}

\begin{figure*}
\includegraphics[height=9 in]{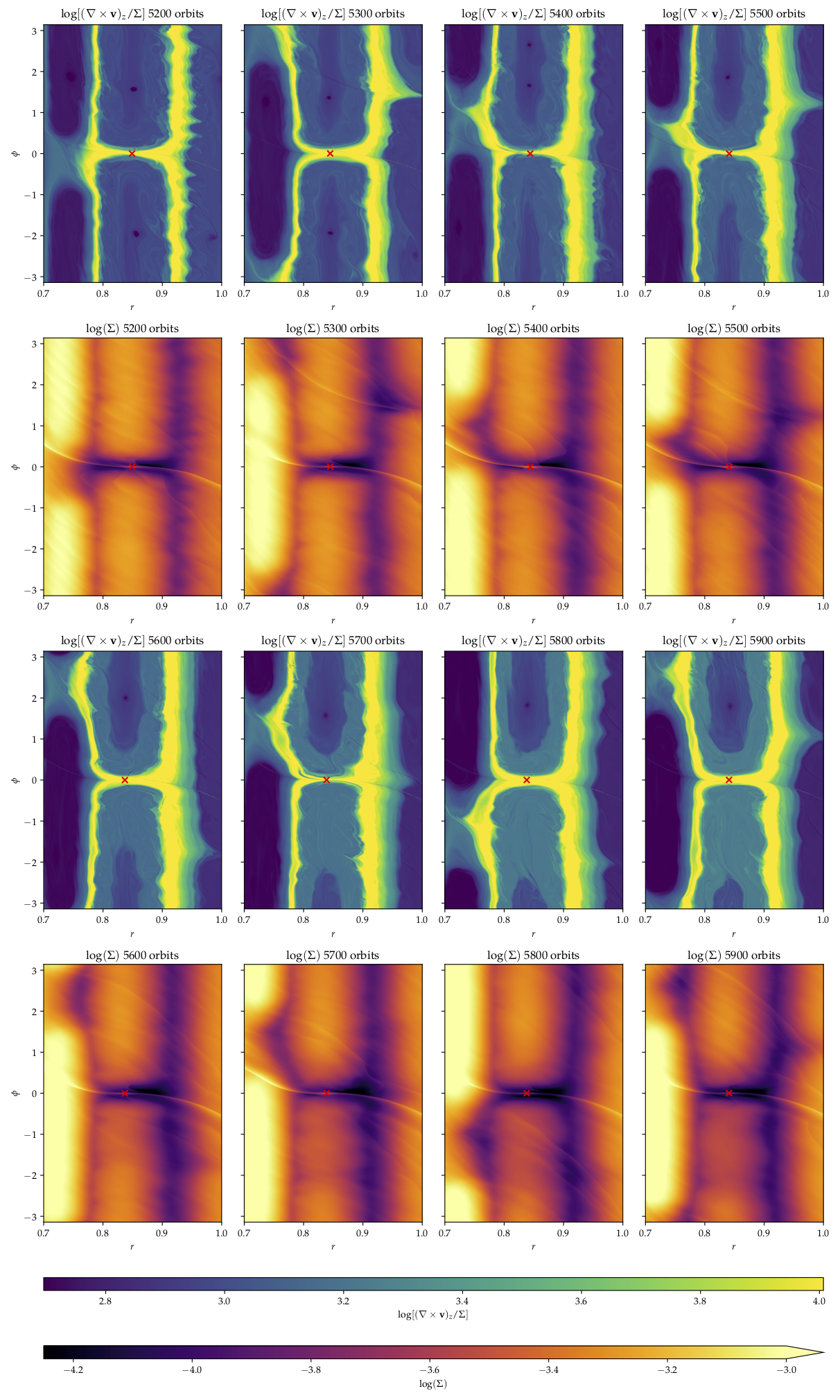}
\caption{Late-time transition from inward to outward migration in the $q=3\times10^{-5}$, r8 inviscid disc run.
The turn in the migration direction is driven by the merger of vortices in the corotation region into a single one sitting in front of the planet. 
By the later stages the single merged vortex drives a significant density asymmetry, producing a positive torque on the planet. 
This can be seen occurring through the panels for 5400, 5500, 5600 orbits}
\label{fig:gap29_sigma_turn}
\end{figure*}

In the highest r8 resolution case of the $q=3\times 10^{-5}$ mass planet, a sustained relatively rapid 
and outward migration episode occurs at late times.
Here, we examine the situation surrounding this, and present the underlying phenomena.
A local view of the planet trajectory around the transition to outward migration is shown in Figure~\ref{fig:gap29_turn}.
The change of migration direction occurs around the period 5600-5700 orbits.
Azimuthally averaged surface density profiles around the planet's location before and after the  
migration reversal are shown in Figure~\ref{fig:gap29r8_profiles}. It is clear from inspection of these profiles that the outward migration episode is \emph{not} being driven by changes to the balance between the inner and outer Lindblad torques. For this to be the case, we would expect the changes to the surface density profiles observed in Figure~\ref{fig:gap29r8_profiles} to be in the opposite sense to what is seen.

We observe that the reversal of the planet's migration direction corresponds to the formation of a single large vortex in the corotation region.
This process is illustrated in the sequence shown in  Figure~\ref{fig:gap29_sigma_turn}. 
While the planet is migrating inwards, two strong vortex cores can be seen in the corotation region, 
notable as the black dots in front of and behind the planet in the 5200 and 5300 orbit 
panels of vortensity in Figure~\ref{fig:gap29_sigma_turn}. 
In the 5400 orbit panel, the two vortex cores are on the same side of the disc as the planet, 
and by 5500 orbits have merged into one vortex.
As this happens, the vortensity and surface density distributions in the corotation region becomes strongly  asymmetrical in azimuth $\phi$.
Through the rest of the panels the single large coorbital vortex tightens, and the surface density distribution 
can be seen to also develop a strong asymmetry in the corotation region.
As the mass asymmetry develops, the planet migration turns around, and past 5700 orbits it is headed outwards.

As indicated above, the behaviour seen here is not the simple feedback-modified Lindblad-torque-based halting 
expected from linear theory.
Additionally, this behaviour is distinct from the runaway Type~III episodes observed in other runs.
The question of whether or not this outward migration can be sustained is uncertain, but it seems unlikely.

 Treating the vortex as a long-lived discrete entity, we might expect that over long time-scales it would horseshoe orbit with the planet, similar to the coorbital Saturnian satellites Janus and Epimetheus, and as such the planet and vortex would periodically swap orbital locations in the gap, limiting the degree of the observed outward migration. Unfortunately, the severe computational demands associated with running this r8 resolution simulation prevent us from undertaking an ensemble study to determine whether or not such horseshoe motion can occur and be sustained, although we note that the outward migration seen so far does not take the planet outside of its existing partial gap.
We further note that as the planet moves outwards, it approaches the outer edge of its partial gap, 
and if the outer Lindblad torque is sufficiently strong the effect of the coorbital vortex may be negated, increasing the complexity of the migration behaviour.
  Indeed in this r8 simulation, run further than shown in Figure~\ref{fig:gap29} the planet migration reverses again and resumes inward motion after approximately 8000 orbits at a radial position $r_{\rm p}\sim 0.96$. 
As with other behaviours, the three dimensional evolution of vorticity may also prove important
 for the long term evolution of these highly non-linear states and needs to be investigated.

\subsection{Low-viscosity discs}
\label{sec:lowvisc}

\begin{figure}
\includegraphics[width=\columnwidth]{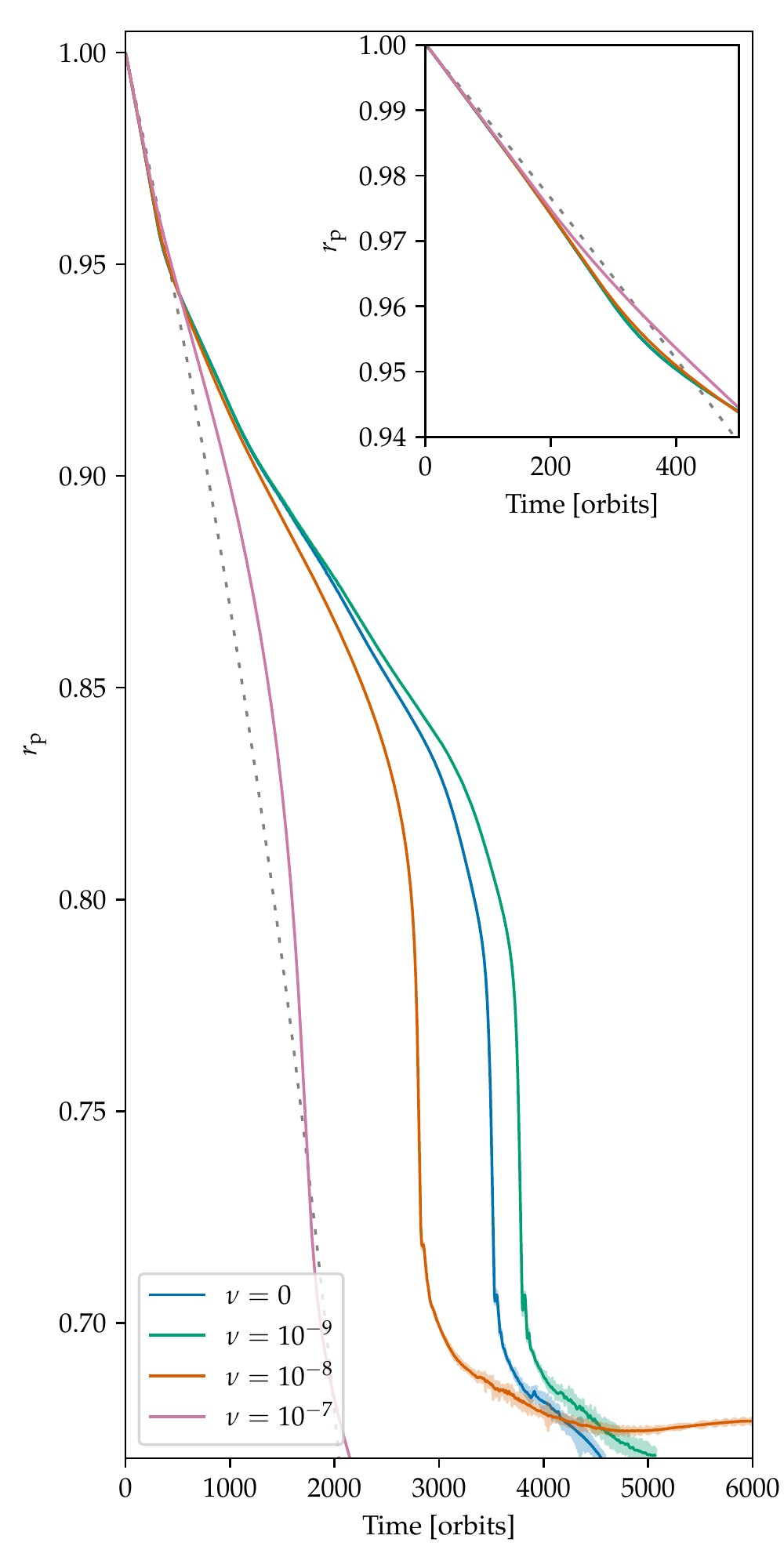}
\caption{$q=1.25\times 10^{-5}$  viscosity study at resolution r4.  The dashed line shows the Type~I migration trajectory.}
\label{fig:1p25visc}
\end{figure}

\begin{figure}
\includegraphics[width=\columnwidth]{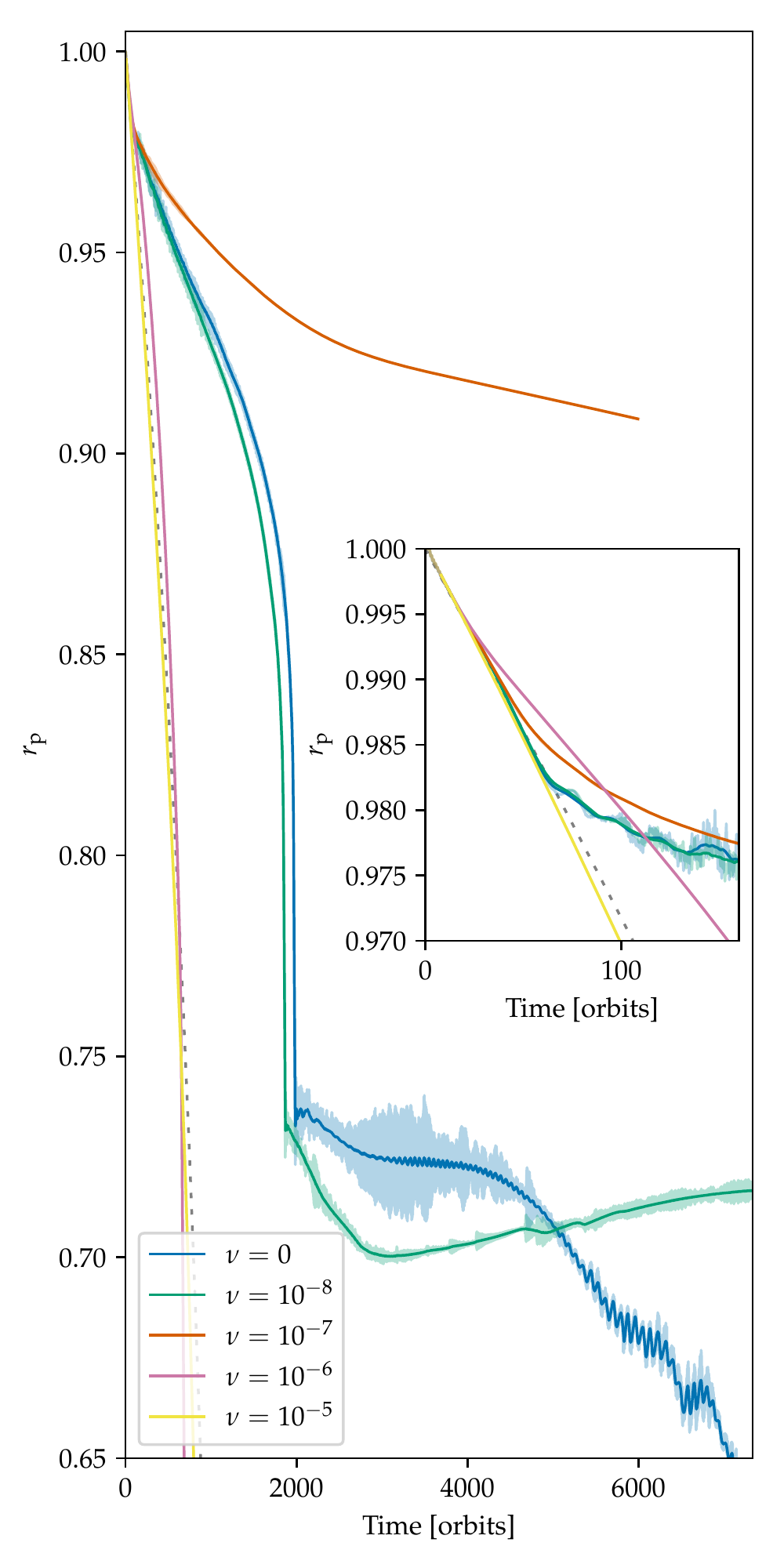}
\caption{ $q=3\times 10^{-5}$  viscosity study at resolution r4.  The dashed line shows the Type~I migration trajectory. }
\label{fig:3visc}
\end{figure}

\begin{figure}
\includegraphics[width=\columnwidth]{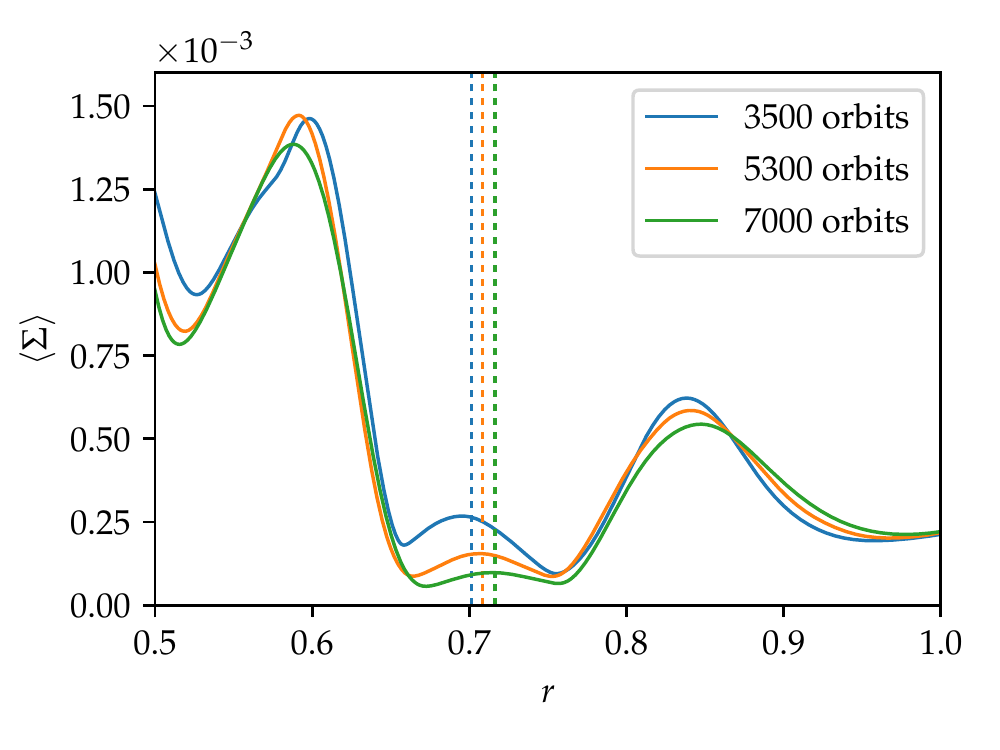}
\caption{$q=3\times 10^{-5}$,  $\nu_0=10^{-8}$ azimuthally averaged surface density
 gap profiles during late-time slow outward migration. Dashed vertical lines mark the instantaneous planet position at the time shown.
 Note the gap bottom moves outwards as the planet moves outwards.}
\label{fig:gap32r4}
\end{figure}

\begin{figure}
\includegraphics[width=\columnwidth]{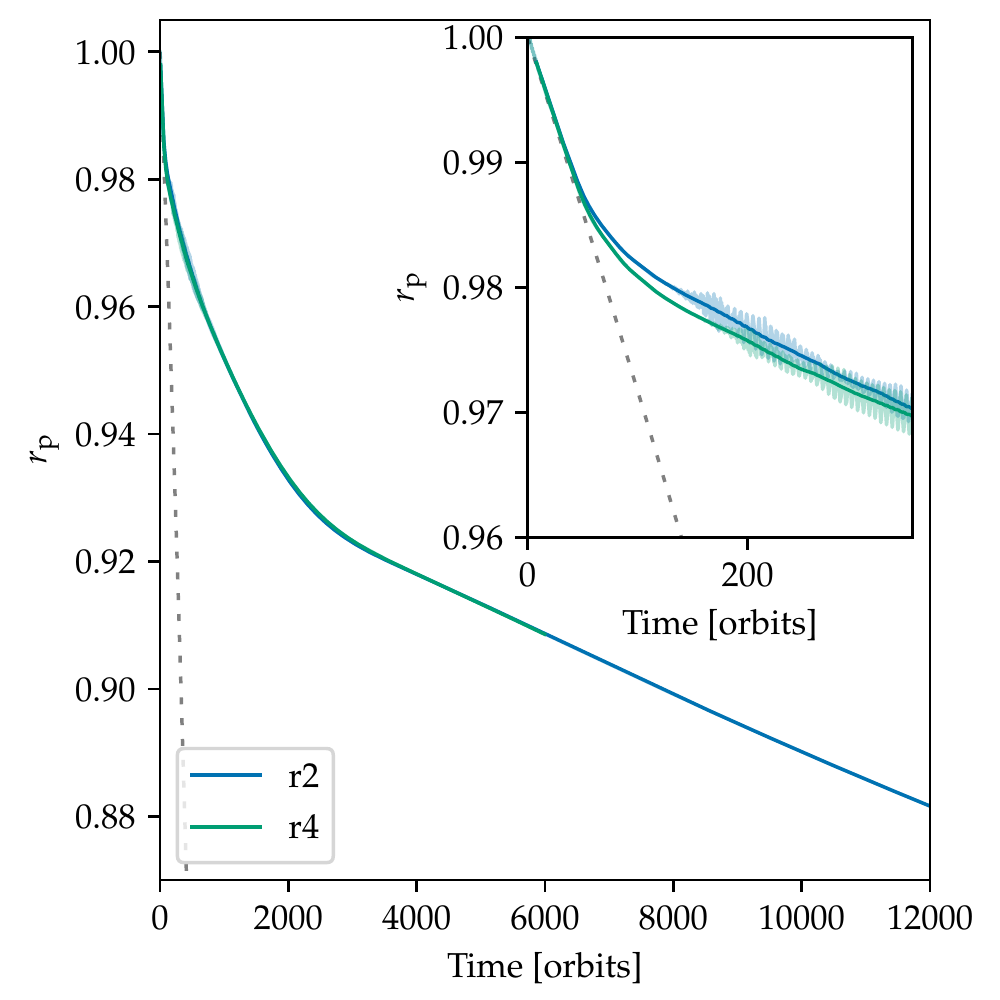}
\caption{$q=3\times 10^{-5}$,  $\nu_0=10^{-7}$ resolution study showing excellent convergence in the long-term planet trajectory, 
although the details of the vortex actions differ on short periods.}
\label{fig:gap31r2}
\end{figure}

\begin{figure}
\includegraphics[width=\columnwidth]{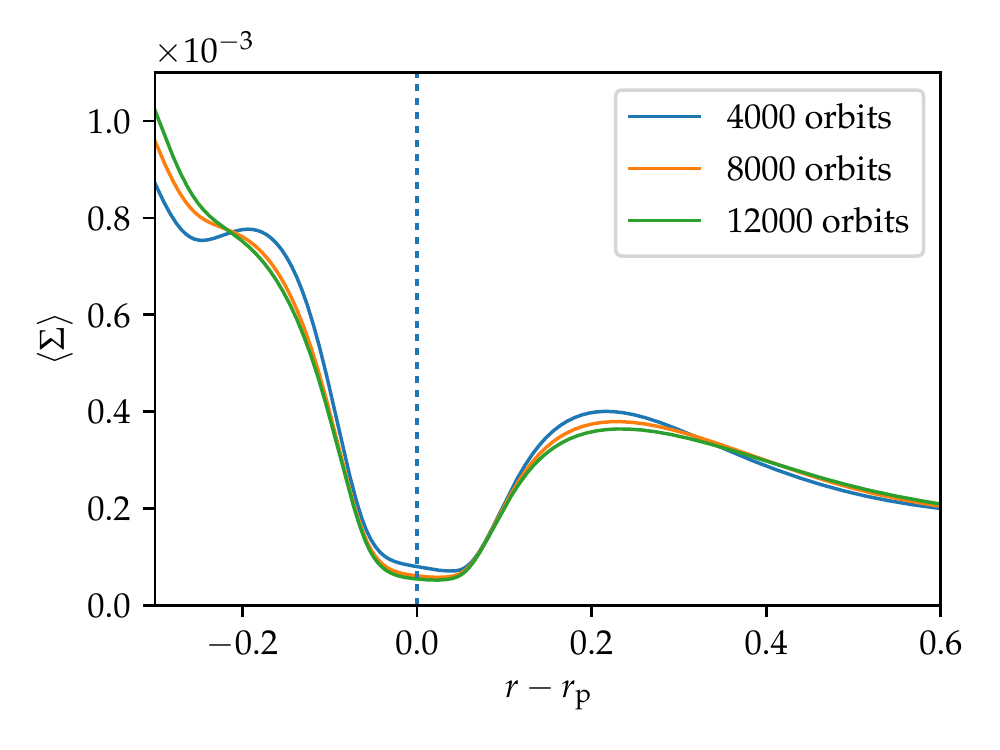}
\caption{ $q=3\times 10^{-5}$,  $\nu_0=10^{-7}$ azimuthally averaged surface density profiles centred on the planet location. 
The planet is located at $\rp$, as denoted by the dashed vertical line. 
The gap profile continues to evolve as the planet slowly migrates on the same time-scale.}
\label{fig:gap31r2_profile}
\end{figure}

The previous section presented numerical simulations in the inviscid limit at a series of different resolutions, 
where only the finite resolution numerical effects at the grid scale lead to diffusion in the flow.
In this section, we present a series of simulations with varying viscosity, so that an explicit finite momentum diffusion is present.
This viscosity can have a range of effects, depending on its magnitude.
Viscosity can either smooth the gap sufficiently to prevent vigorous RWI and vortex formation, 
making feedback based slowing of migration more effective;
or smooth the surface density features so much that feedback is not effective.

The viscosity is specified with a radial dependence such that it does not transport angular momentum or induce a global radial flow of gas in the initial disc:
\begin{align}
\nu(r) = \nu_0 \left(\frac{r}{r_0}\right)^{\alpha-1/2}\ ,
\end{align}
where $\alpha$ is the disc initial power-law density slope and $\nu_0$ is a constant. 
This viscosity will however induce local radial flows where the planet modifies the surface density of the disc, acting to smooth it.

For the $q=1.25\times 10^{-5}$ ($4.2\ \mathrm{M_\oplus}$) mass planet we find that $\nu_0=10^{-9}$ ($\alphaSS=8\times10^{-7}$) results 
in near-identical results to our r4 resolution inviscid model, and $\nu_0=10^{-7}$ ($\alphaSS=8\times10^{-5}$) 
results in a constant inward migration recognizable as Type~I phenomenology (see Figure~\ref{fig:1p25visc}).
Thus, there is not much room to fine-tune a viscosity which might result in smooth slowing of migration.

At the higher planet mass, $q=3\times 10^{-5}$ ($10\ \mathrm{M_\oplus}$), a similar scan of viscosities at resolution r4 is shown in Figure~\ref{fig:3visc}.
A diverse range of migration behaviours is displayed at the various viscosity values.
At $\nu_0=10^{-8}$ the behaviour is similar to the inviscid run, first undergoing a feedback slowing, growth of vortices, and then an inward Type~III episode.
From 3500 orbits onwards, it experiences a slow outward migration, which will be discussed below in Section~\ref{sec:slowmigpartgap}.
At $\nu_0=10^{-7}$ the feedback settles after roughly 3000 orbits in a sustained, slow, inward migration, again discussed  in Section~\ref{sec:slowmigpartgap}.
At $\nu_0=10^{-6}$ the initial feedback  slowing does not occur, and instead the planet immediately enters Type~III inward migration and rapidly migrates inwards.
At $\nu_0=10^{-5}$ the viscosity is sufficiently strong to keep the planet in Type~I migration.
Thus, Fig.~\ref{fig:3visc} contains a complete scan across the feedback and vortices, smooth feedback, and 
viscous Type~I migration regimes denoted in the diagram in Figure~\ref{fig:migration_org_chart}.

\subsubsection{Slow migration in a partial gap}
\label{sec:slowmigpartgap}
In the simulation with $q=3\times 10^{-5}$  and $\nu_0=10^{-8}$, the planet undergoes a slow, 
outward radial migration after roughly 3000 orbits.
This is a more pronounced version of the same phenomenon seen in the $q=1.25\times 10^{-5}$, $\nu_0=10^{-8}$
simulation (Figure~\ref{fig:1p25visc}) and in the similar run of \citet{2010ApJ...712..198Y}.
For this particularly clean example, we plot the evolution of the azimuthally averaged surface density in
Figure~\ref{fig:gap32r4}.
The gap profile can be seen evolving, under the joint action of the wake, vortices, and viscosity. 
Notably, the bottom of the gap moves inwards as the planet moves outwards. 
Thus, the outward migration appears to be driven by the relaxation of the gap profile in which the planet sits.

For the $q=3\times 10^{-5}$ ($10\ \mathrm{M_\oplus}$) mass planet, $\nu_0=10^{-7}$ ($\alphaSS=8\times10^{-5}$)  
appears to smooth the surface density and sufficiently damp the RWI vortices to allow feedback-based slowing of planet migration (Figure~\ref{fig:3visc}).
Figure~\ref{fig:gap31r2} demonstrates the excellent convergence obtained in the planet trajectory for this run between the r2 and r4 resolutions.
The evolution of the gap profile during the long smooth-feedback migration phase is shown in Figure~\ref{fig:gap31r2_profile}, 
where, past 400 orbits, the migration time-scale is of3 the order of $2\times 10^5$ orbits.
Our choice of radial scaling of viscosity does not induce a global viscous accretion inflow through the unperturbed disc, 
so this planet migration rate cannot be identified with a background viscous disc inflow rate as it was in \citet{2010ApJ...712..198Y}.

In the interpretation of Type~II migration by \citet{2018ApJ...861..140K}, the migration rate should be 
given by the Lindblad-torque-driven rate associated with the surface density at the bottom of the gap.
However, despite the migration rate from 400 to 12000 orbits being very close to constant, 
the surface density inside the gap varies by $\sim 20$ per cent.
Moreover, taking the surface density in the gap at 4000 orbits, the migration time-scale in 
the \citet{2018ApJ...861..140K} interpretation is only $1.5\times10^4$ orbits, clearly much shorter than the migration time of $\sim 2\times10^{5}$ orbits shown in our simulation .

From the evolution of the gap profile in Figure~\ref{fig:gap31r2_profile}, it appears that 
the relevant time-scale is instead the viscous time-scale for the gap features (note that the viscosity supresses vortices in this run).
Here, for features of a radial width of approximately $0.4$ subject to a diffusion 
on the order of $10^{-7}$ the diffusion time-scale is on the same order as the migration rate.
That the Type~II migration rate is determined by the diffusive relaxation of the gap structures
has been suggested by \citet{2015A&A...574A..52D} and directly investigated by
\citet{2018A&A...617A..98R}.
Here, we find that the smooth feedback regime appears to behave in this way.
In this paradigm, it is possible to interpret smooth-feedback as being simply an extreme of Type II migration.

\subsection{Inviscid discs with a vortensity gradient}
\label{sec:vortensitygradient}

\begin{figure}
\includegraphics[width=\columnwidth]{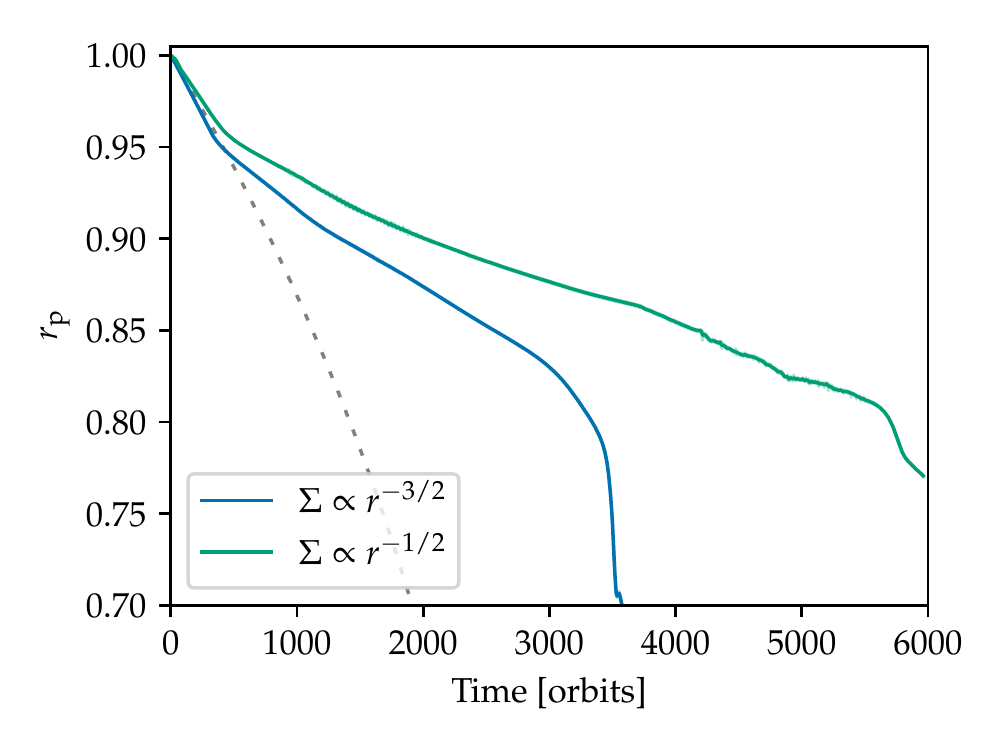}
\caption{$q=1.25\times 10^{-5}$, with a run $\Sigma\propto r^{-1/2}$ compared 
to the equivalent simulation with no vortensity gradient ($\Sigma\propto r^{-3/2}$).
The planet momentarily slows at 4200 orbits, but renewed vortex action speeds its inward migration.
The dashed line shows the Type~I migration trajectory for the $\Sigma\propto r^{-3/2}$ case. }
\label{fig:gap43}
\end{figure}

\begin{figure}
\includegraphics[width=\columnwidth]{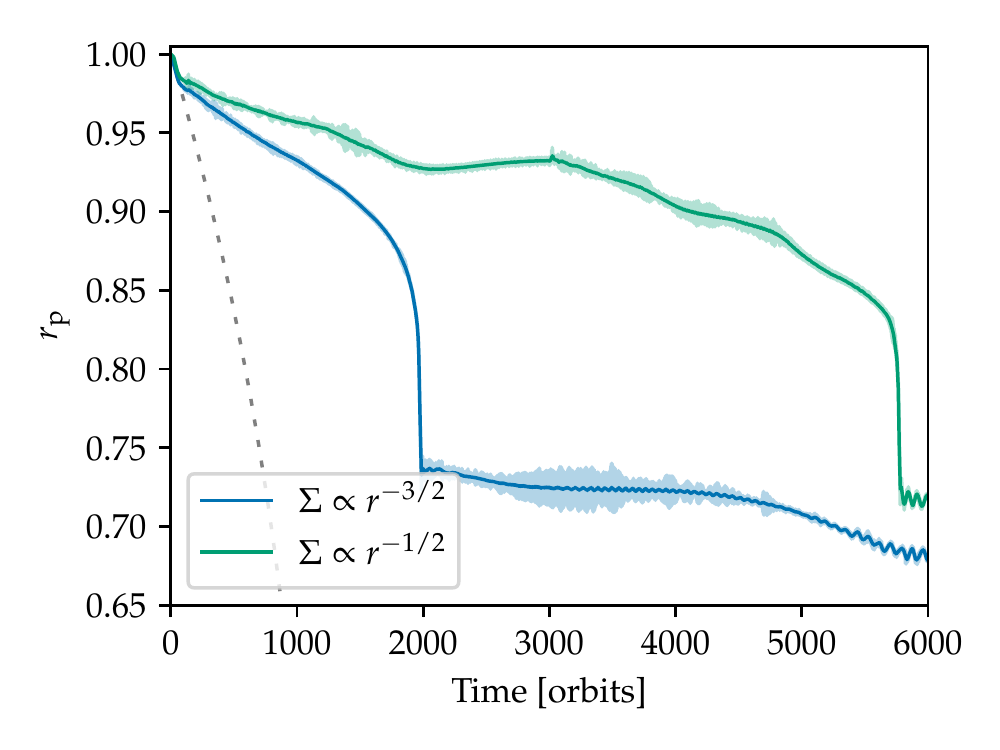}
\caption{$q=3\times 10^{-5}$, inviscid model with $\Sigma\propto r^{-1/2}$ compared 
to the equivalent simulation with no vortensity gradient ($\Sigma\propto r^{-3/2}$).
The dashed line shows the Type~I migration trajectory for the $\Sigma\propto r^{-3/2}$ case.}
\label{fig:gap40}
\end{figure}

Previous studies have not been undertaken to study the effect of a vortensity gradient on feedback modified migration.
Here, we introduce a vortensity gradient to the models run in Section~\ref{sec:inviscid}, and compare the planet migration behaviour.
The vortensity gradient is introduced by altering the radial power law slope of the surface density in the disc model in Section~\ref{sec:inviscid}, 
while retaining the same globally isothermal structure.
First, a planet with mass $q=1.25\times 10^{-5}$ in an inviscid disc with surface density law $\Sigma\propto r^{-1/2}$ 
at resolution r4 is shown in Figure~\ref{fig:gap43} along with its isovortensity partner.
Most apparent in the migration trajectory is that the introduction of the vortensity gradient slows the migration, both before 
and after the disc feedback effect becomes strong at around 400 orbits.
That this should be the case is somewhat surprising, as once the planet has altered the surface density enough to change the migration rate the connection with the disc global gradients is inherently weakened.
Additionally, the tendency of the planet to undergo Type~III migration appears to be lessened, however it does undergo a short episode at  $\sim 5700$ orbits.
For the larger planet with mass $q=3\times 10^{-5}$, the results of the same experiment in an inviscid disc at resolution r4 with surface density law $\Sigma\propto r^{-1/2}$ are shown in Figure~\ref{fig:gap40}.
Again, from the beginning of the evolution the planet migrates more slowly with the vortensity gradient than without.
The time-averaged torque density profiles for these two runs are presented in appendix~\ref{app:vorttorque}.
The planet is even able to turn around and migrate outwards from roughly 2000 to 3000 orbits, before a burst of vortex activity reinitiates inward migration.
Here too, the tendency to undergo Type~III episodes appears to be lessened, with one eventually occurring at roughly 5700 orbits.
Notably, the Type~III episode terminates at roughly the same radial location in the  $\Sigma\propto r^{-1/2}$ run as in the  $\Sigma\propto r^{-3/2}$ one.
This location corresponds to the position of the density peak on the inside of the initial partial gap carved by the planet.

\subsection{Advective discs}
\label{sec:advectivedisc}

\begin{figure}
\includegraphics[width=\columnwidth]{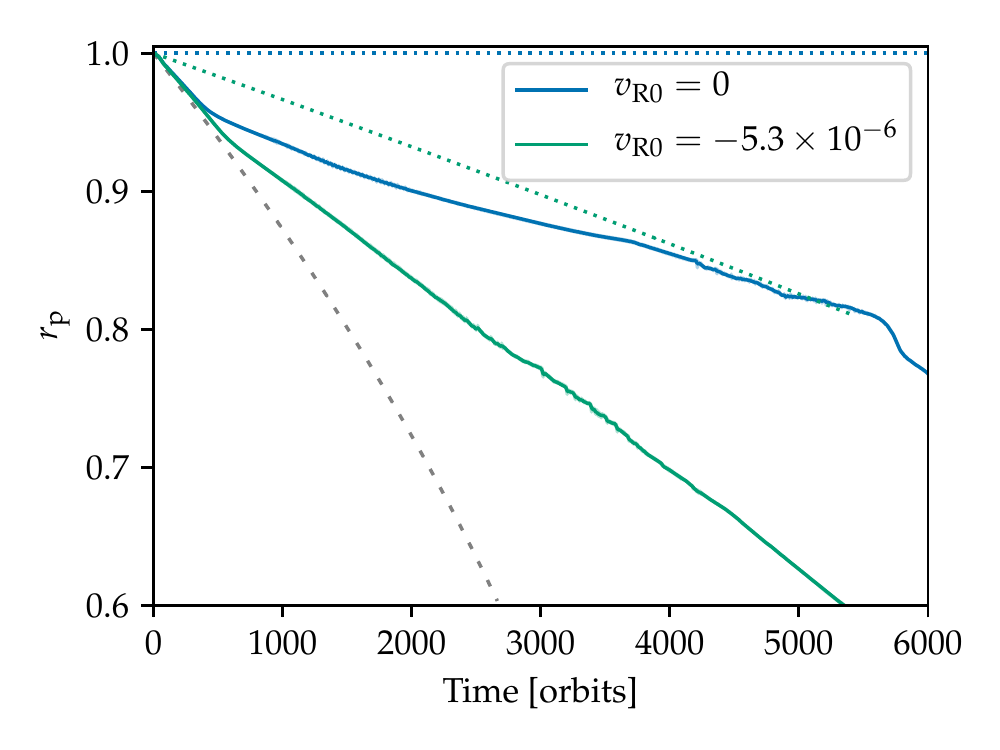}
\caption{$q=1.25\times 10^{-5}$, with a run $\Sigma\propto r^{-1/2}$ and an advective disc inflow. 
Disc gas inflow trajectories are shown by dotted lines, and Type~I migration is shown by the dashed line  (assuming saturated corotation torque).}
\label{fig:gap45}
\end{figure}

\begin{figure}
\includegraphics[width=\columnwidth]{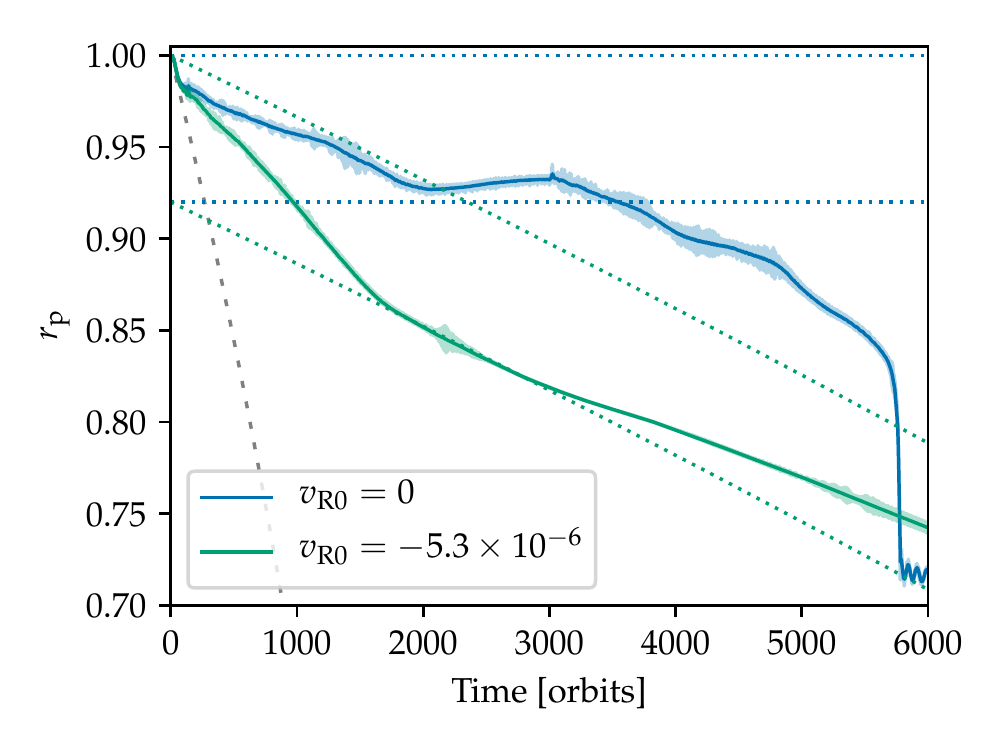}
\caption{ $q=3\times 10^{-5}$,  in an inviscid disc with  $\Sigma\propto r^{-1/2}$  and resolution r4. 
Runs with an advective disc inflow ($v_\mathrm{R0}=-5.3\times10^{-6}$) and without ($v_\mathrm{R0}=0$). 
Disc gas inflow trajectories shown by dotted lines, and Type~I migration in a static disc shown by the dashed line (assuming saturated corotation torque).}
\label{fig:gap41}
\end{figure}

In this section we extend the study of inviscid feedback-modified migration to the `advective disc' 
model introduced in \citet{2017MNRAS.472.1565M} and \citet{2018MNRAS.477.4596M}.
The simulations are a modification of those in Section~\ref{sec:vortensitygradient}, with an added torque causing the disc gas to flow inwards with a spatially constant mass flow rate. As discussed in Section~\ref{sec:ConnectingModels}, the inviscid disc models without accretion flows presented so far in this paper apply to magnetized disc models where only a very small column density of gas is coupled to the magnetic fields, typically expected when the vertical magnetic field and disc rotation vectors are anti-aligned. The models in this section apply to the aligned configuration, where the accretion flow can involve a large fraction of the disc column density.
 
Given the planet mass and surface density parameters studied here, the planet's inward migration velocity under Lindblad torques alone in the unperturbed disc is faster than any reasonable disc inflow velocity.
For example, the Lindblad-torque-driven Type~I migration rate of a planet more massive than $q=8\times10^{-7}$ ($0.27~\mathrm{M_\oplus}$) 
exceeds the gas inflow velocity of an advective version of the  2~MMSN--like disc used here if the disc accretion rate is $\dot{M}=10^{-8}~\mathrm{M_\odot\ yr^{-1}}$.
Thus, we only probe the case where the planet is initially migrating inwards faster than the disc gas.
This is the configuration of planet and disc velocities described in \citet{2017MNRAS.472.1565M} as being in their regime~(i).
The first simulation with the planet with mass
$q=1.25\times 10^{-5}$ and gas inflow speed of $v_\mathrm{R0}=-5.3\times10^{-6}$ is shown in Figure~\ref{fig:gap45}.
The planet does not asymptotically lock to the disc gas inflow velocity as was observed in \citet{2017MNRAS.472.1565M,2018MNRAS.477.4596M}, who considered planet masses below the feedback mass $M_{\rm F}$.
Unlike in those lower mass cases, here gap formation and strong vortex action prevents the corotation torque from achieving this locking. However, compared to the equivalent inviscid simulation conducted without a laminar accretion flow (discussed in Section~\ref{sec:inviscid}), we do observe that the migration velocity is boosted by approximately the disc inward flow velocity. That is, the migration rate is constant with respect to the disc gas, having the same sort of Galilean invariance as was found for low-mass planets in 
\citet{2017MNRAS.472.1565M, 2018MNRAS.477.4596M}.
The planet trajectory from a simulation with the same advective disc setup but with the higher mass planet $q=3\times 10^{-5}$ is shown in Figure~\ref{fig:gap41}.
Once disc feedback begins to have a significant effect, the planet motion with respect to 
the disc gas is roughly the same in the cases with and without gas inflow up until roughly 3000 orbits.
Here, the planet does end up with a migration motion effectively locked to the disc inflow, at times between approximately 1700 and 2700 orbits. This roughly corresponds to the period when the planet was migrating very slowly in the same setup without disc inflow. However, for this larger planet mass, as the gap profile evolves the planet migrates inwards more slowly than the disc at later times. 
As a result, gas slowly piles up outside the gap, and the gap becomes increasingly asymmetrical as the inner edge is advected inwards away from the planet.

\subsection{Giant planets in inviscid discs}
\label{sec:typeII}

\begin{figure}
\includegraphics[width=\columnwidth]{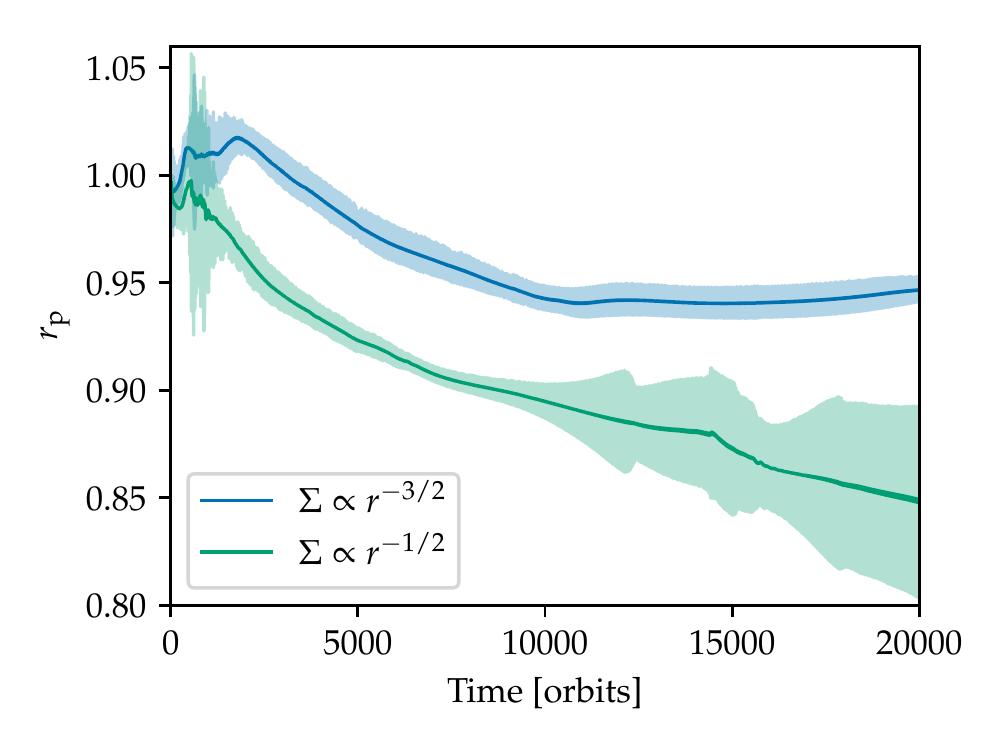}
\caption{$q=3\times 10^{-4}$, with two surface density profiles. Type II with vortices. }
\label{fig:typeII}
\end{figure}

We now consider the migration of significantly more massive planets than discussed above in inviscid discs without accretion flows, placing the planet unambiguously in the Type~II migration regime.
Here, vortices have a very strong tendency to form at the gap edges in an inviscid disc \citep{2003ApJ...596L..91K,2007A&A...471.1043D}.
Jupiter at 5.2 au in an  MMSN is a canonical example of a planet unambiguously in the Type~II migration regime, 
so the relevant scalings dictated by the non-linearity of the disc response to the planet have been applied here to choose the planet mass. Scaling the planet mass by $h^3$, a planet mass $q=3\times10^{-4}$, is the equivalent in the $h=0.035$ disc to a Jupiter-mass planet in a disc where $h=0.05$, thus the experiment used this value.

As the planet was massive enough to accrete a massive bound atmosphere inside the Hill region, the 
appropriate torque calculation for the planet motion was different in these simulations than in the previous ones with lower mass planets. In these models, the torque from the material in the planet's Hill sphere was removed from the calculation of the migration torque in a manner similar to \citet{2008A&A...483..325C}. 
 For a Hill sphere with radius $r_{\rm H} = (q/3)^{1/3}r_{\rm p}$, all material within $1/2 r_{\rm H}$ was neglected from the gravitational force calculation on the planet, and the contribution from cells at distances $s$ between $1/2 r_{\rm H}$ and $r_{\rm H}$ was reduced by a factor $\sin^2((s/r_{\rm H}-1/2)\pi)$. \footnote{ This is enabled by the {\sc HILLCUT} option in {\sc FARGO3D} 1.3.} 
Qualitatively, this high-mass planet opens a very wide and deep gap, and represents a clear transition across the gap opening line in Figure~\ref{fig:migration_org_chart}.
The simulations were performed on the r2 grid with the same initialization as previous runs.
Figure~\ref{fig:typeII} shows the planet trajectories, for runs with a disc surface density gradient $\Sigma \propto r^{-3/2}$ and $\Sigma \propto r^{-1/2}$.

It is immediately apparent that after an initial period of gap formation and surface density readjustment, vortices are continually driven in these inviscid discs, and the planet experiences a slow inward migration.
This is in contrast to the classical Type~II migration concept where the planet migration is locked to the disc's viscously driven flow, which is not present in this inviscid disc.
A second observation  that can be drawn from this experiment is that the despite differing initial behaviour in the first $\sim 2000$ orbits, the inward migration does not show a strong dependence on the disc surface density gradient during intermediate times between $\sim 2000$ and $\sim 6000$ orbits.
This stands in contrast to the behaviour of lower mass planets in Section~\ref{sec:vortensitygradient}, where the disc vortensity and surface density gradients had a significant effect on the planet's migration rate.

In the viscous context, competing theories have been proposed to explain Type~II migration proceeding faster than the disc viscous inflow rate.
Variously, the migration is attributed to mass crossing the gap, the torque on the planet exerted by the remnant gas inside the gap, 
or the shifting of the minimum torque position by viscous smoothing of the gap edge structures.
In the case of an inviscid disc, with tidal truncation of the gap preventing significant gas flows across the planetary orbit, we would predict that smoothing of the gap edges by vortices would induce only a transient phase of Type II migration, with the migration stalling after the planet has moved a distance similar to the gap width. The logic here is simply that the gap opening planet displaces material from the gap into the pressure bumps at the edge of the gap. The resulting formation of vortices via the RWI, and their smoothing effect on the gap edges, induces the planet to move inwards until the displaced material and pressure bumps have been smoothed back towards their original state, after which migration essentially stalls as the planet loses contact with the outer gap edge. Interestingly, at late times the simulations seem to support this picture, with the Type II migration stalling after $\sim 11000$ orbits. However, the simulation with surface density $\Sigma \propto r^{-1/2}$ also shows that bursts of gap edge vortices formation may be able to diffuse material through the gap, such that migration can be sustained, albeit in a somewhat erratic stop--start fashion.

\subsection{Pebble isolation due to pressure maxima}
\label{sec:pebbletrap}

\begin{figure*}
\includegraphics[width=1.2\columnwidth]{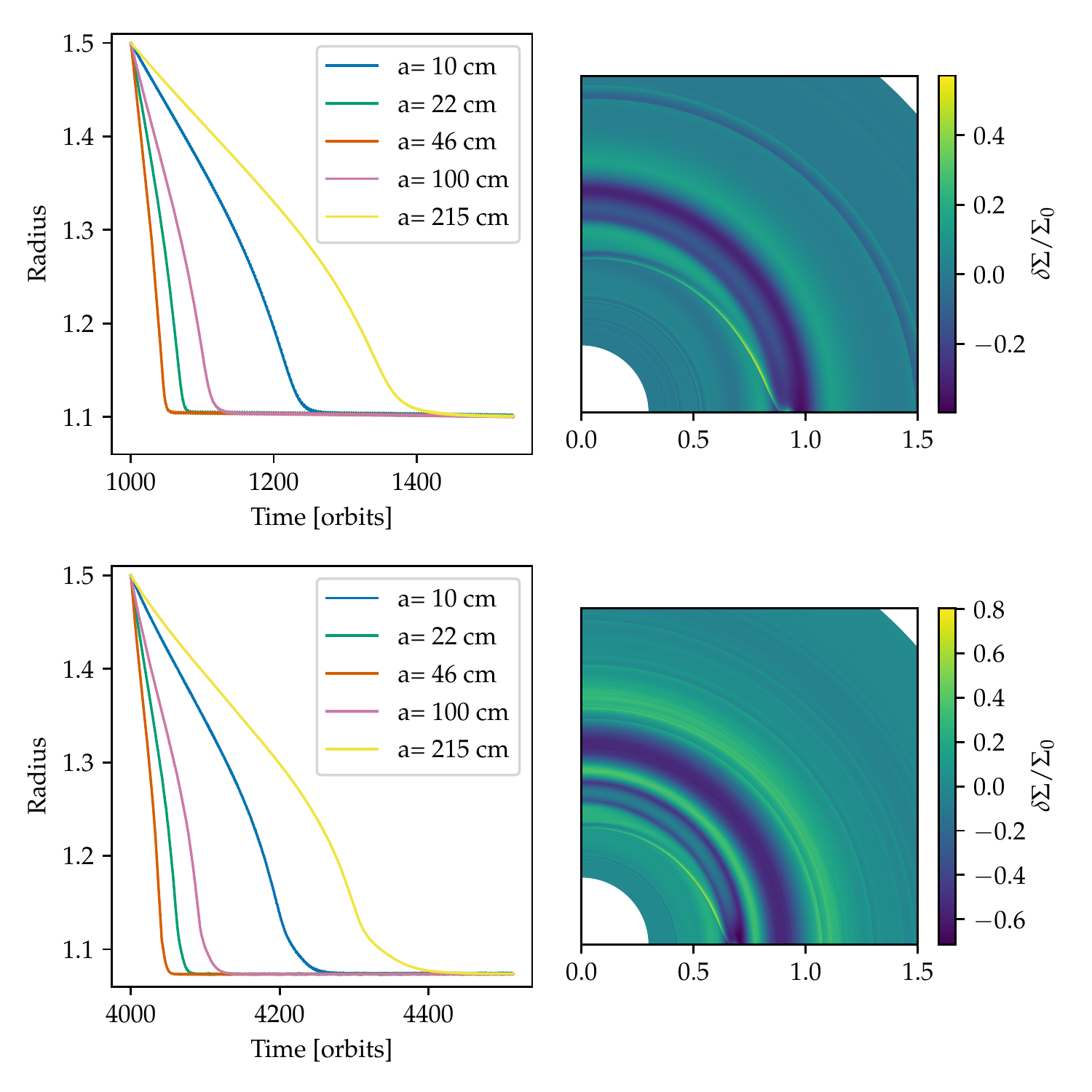}
\caption{$q=1.25\times 10^{-5}$ ($4.2\ \mathrm{M_\oplus}$)  particle trajectories, 
with start times of 1000 orbits (top panels) and 4000 orbits (lower panels), before and after the rapid inward migration episode. Left column: particle trajectories, 
by particle radius $a$ showing particle trapping in the outermost pressure bump. 
Right column: disc surface density relative perturbation, showing one quadrant of the disc. }
\label{fig:pebbletrap}
\end{figure*}

As soon as a planet in an inviscid disc is able to form a gap and generate a local pressure maximum it can produce a blockage of the radial drift of solids through the disc, creating a pebble trap \citep{2014A&A...572A..35L}. In the context of viscous discs, the planet mass for which this occurs (the so-called pebble isolation mass) is typically $\sim 20\ \mathrm{M_\oplus}$, but depends on disc parameters \citep{2018A&A...612A..30B}. 

Formally speaking, in an inviscid disc, if even a very low mass planet is held at the same position with respect to the disc gas for long enough,
it will eventually generate a pebble trap as a consequence of the non-local wake dissipation gap opening process described by \citet{2002ApJ...569..997R}. Above the feedback mass, a freely migrating planet will more easily be able to generate a sufficient modification to the disc surface density to form a pebble trap.
Here, we demonstrate that pebble traps are formed by the lowest mass planet used in this study. 
 
In Figure~\ref{fig:pebbletrap} the trajectories of solid particles introduced into the simulation with a $q=1.25\times 10^{-5}$ ($4.2\ \mathrm{M_\oplus}$) planet in an inviscid disc at resolution r4 (described in Section~\ref{sec:inviscid}) are shown. These passive particles of varying radii $a$ are introduced at the two different times shown at a radius $r=1.5$, and feel a drag force from the disc gas, gravitational force from the planet, and the central star.
Particles were taken as silicate spheres with a density of $3\ \mathrm{g\ cm^{-3}}$ with the drag law as used by \citet{2010MNRAS.409..639N}, for a $5:1$ mixture of H$_2$ to He by number.
The trajectory integration scheme used was a kick--drift--kick variation of that described for the local simulations in \citet{2010MNRAS.409..639N}, so that the symplectic property is recovered for loosely coupled particles.

The initial batch of particles were introduced to the model after 1000 orbits, 
while the planet's migration has been slowed by disc feedback, but it has not yet undergone a Type~III episode (Figure~\ref{fig:gap16}, resolution r4). Hence, the particles represent pebbles drifting in from disc regions close to the planet's putative formation location.
Trapping of the solids occurs near $r=1.1$, in the pressure and density bump created by the planet's wake to the outside of its orbit, with the planet at a radial position of approximately $0.9$ and moving inwards.
A second batch of particles were introduced into the model at the same radial location much later at 4000 orbits. Hence, these represent pebbles migrating inwards from further out in the disc and arriving at later times. Surprisingly, although the planet is now at a very different radial location, around $r=0.65$, the particles are trapped at roughly the same radial location as before at $r=1.1$.
Even after the planet has undergone an episode of rapid Type~III inward migration, the pressure bump left by the 
wake dissipation outside its former location in the disc persists and acts as a very effective pebble trap. 
The pressure bump at $r\sim1.1$ is long lasting as there no viscosity to smooth it.
At this later time, there is a pressure maximum formed immediately outside the planet's location, at $r\sim0.75$, but pebbles from the outer disc are trapped instead by the pressure bump left at $r\sim1.1$ and never reach this inner pebble trap. 
 Similar pebbles will continue to be captured by the outer pressure bump until it is dissipated, 
by for example hydrodynamic processes or momentum exchange with the inflowing solids. 
Thus, in an inviscid disc, a single, relatively low mass planet can cut off the pebble flux from a large radial domain, even after it has migrated inwards. 
This has clear consequences for planetary system formation via pebble accretion. In particular it has been suggested that the formation of Jupiter early in the history of the Solar system prevented a flux of pebbles reaching the inner terrestrial planet system, hence explaining the dichotomy between low-mass planets in the inner Solar system, and giant planets further out \citep{2015Icar..258..418M}. If the solar nebula was low-viscosity, as suggested by modern models of magnetized discs, then the mass at which proto-Jupiter was able to cut off the pebble flux was significantly smaller than suggested by previous estimates \citep[see also][]{2018ApJ...859..126F}.

\subsection{Multiple dust rings from a migrating planet}
\label{sec:yarfm}

\begin{figure}
\includegraphics[width=\columnwidth]{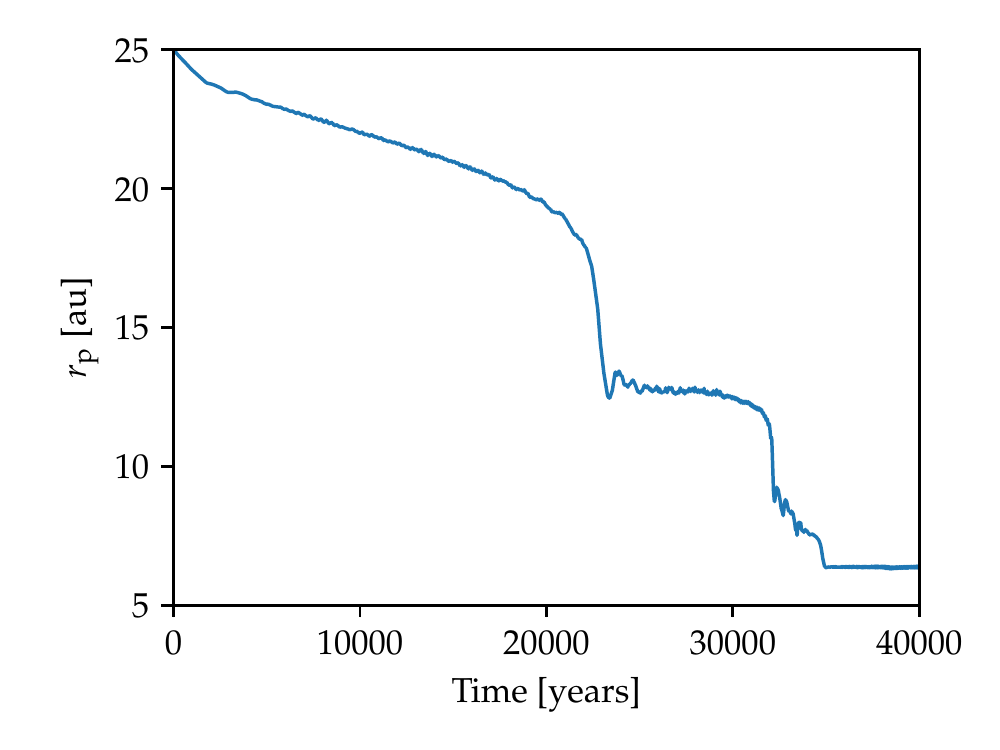}
\caption{Planet undergoing repeated Type~III migration episodes from $25$~au in a 2 MMSN--like disc.}
\label{fig:inrings1rad}
\end{figure}

\begin{figure}
\includegraphics[width=\columnwidth]{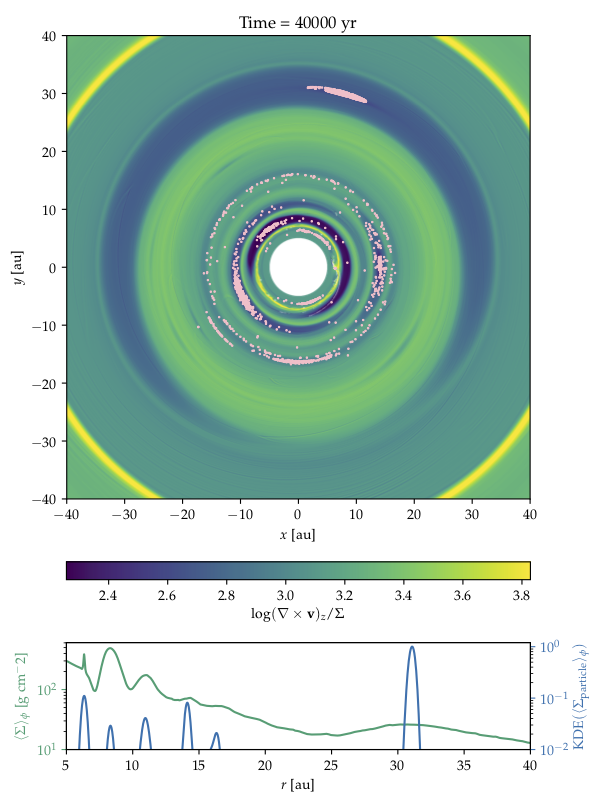}
\caption{{\sl Upper panel}: Logarithm of vortensity, 
showing vortices and rings of enhanced vortensity left in the inviscid disc by the passage of the planet, 
with positions of $1$~mm silicate particles allowed to drift in the gas.
The planet has migrated from its starting position at $25$--$6.4$~au 
where its migration has been halted by interaction with the inner boundary wave damping zone.
The yellow ring is an numerical artefact which arises where the grid spacing function changes.
{\sl Lower Panel}: Azimuthally averaged profiles of the gas surface density, 
and of the Gaussian kernel-density-estimate (bandwidth $0.02$~au) of the solid particle density, normalized to the peak value.}
\label{fig:inrings1part}
\end{figure}

As shown in the previous section, features induced by the planet's wake can persist long after the planet has migrated inwards.
In particular, if the planet undergoes rapid Type~III inward migration episodes.
This scenario is particularly enabled if the planet reaches the feedback mass at a large radius in the disc, as it may then have the 
opportunity to undergo multiple rapid inward migration episodes, at radial scales observable to current submillimetre interferometers such as ALMA. 
Here, we present such a model, where we also trace the evolution of a population of dust particles which would be similar to those observable
to show how they are concentrated by the vortices and rings left behind by the passing planet.

In this section, we define the model in physical units with a central stellar mass of $1\ \mathrm{M_\odot}$, 
giving lengths in astronomical units and times in years.
A  planet with mass ratio $q=4\times10^{-4}$ ($0.42$ Jupiter masses) was placed initially at a radius $25$~au in a 2 MMSN inviscid disc based on that in Section~\ref{sec:scales}, 
again here globally isothermal, but with the temperature set so that the aspect ratio $h=0.08$ at $25$~au, accounting for the flared structure of the disc.
The mass is chosen so that it is $1.5\ M_\mathrm{F}$, the same relative to the feedback mass as the lower 
mass planet used in Section~\ref{sec:inviscid}.
In addition, 20000 $1$-mm radius silicate particles were introduced to the disc, initially on 200 radial rays, 
with 100 particles spaced evenly between $8$ and $60$~au, giving a surface density of dust particles initially $\propto r^{-1}$.
As the particles only drift due to gas drag and gravity, and do not diffuse or collide and fragment, 
these only trace the concentration of dust produced by the vortices and pressure bumps in the disc. 
The slope of this initial distribution serves mainly to bias how many particles will be available to 
highlight each area of concentration in the final result.

Figure~\ref{fig:inrings1rad} shows the migration trajectory of the planet.
After release, the planet undergoes the now familiar feedback slowing, followed by an episode of rapid 
Type~III migration from roughly $19$ to $13$~au. 
After several oscillations, it undergoes another rapid inward migration episode to roughly $9$~au.
It then undergoes a slightly more complicated series of oscillations and inward migration, followed by a clear 
Type~III episode from roughly $7.5$~au until its migration is halted by interaction with the inner boundary damping zone of the grid.
Similarly to the result shown in Section~\ref{sec:pebbletrap}, the result of the Type~III episodes is to leave structures 
induced by the planet's wake in the disc far outside of the eventual location of the planet. 
Repeated Type III migration, driven by interaction with vortices, has previously been examined in depth by \citet{2010MNRAS.405.1473L}. 
Figure~\ref{fig:inrings1part} shows the vortensity of the disc at $40$~kyr, highlighting the vortices and rings of differing vortensity left in the disc by the passing planet.
Overplotted in pink points are the positions of the $1$~mm silicate particles which have been allowed to drift in the gas, 
collecting in the high pressure regions, corresponding to rings of low-vortensity and azimuthally confined vortices.
As the only source of vortensity non-conservation in this barotropic flow is the curved shock of the planetary wake, 
it can be clearly seen that the particles are concentrating at structures created by this wake, 
even long after the planet has migrated to smaller radii.
The particles collecting in these features are physically on the scale of those detectable 
in submillimetre observations of protoplanetary discs, such as those obtained by ALMA, 
where the features consist of a combination of rings and vortices - a situation which is probably 
unique to inviscid or very low-viscosity discs.\footnote{
During preparation of this manuscript, we became aware that 
Gaylor Wafflard-Fernandez and Clement Baruteau have coincidentally 
discovered a similar scenario for migration with multiple Type~III phases and resulting pattern formation in the dust.} 
Hence, the types of features observed in protoplanetary discs may provide important constraints on the levels 
of turbulence present since rings but not vortices may be present in discs that sustain significant levels of turbulent viscosity.
Recent observations of HD~135344B by \citet{2018A&A...619A.161C} show a tantalizingly similar structure 
in the dust to that found here, with an apparent dust-trapping vortex immediately outside of a dust ring.

Recent studies have considered other cases where a single planet may cause multiple ring
 features to occur in a low-viscosity disc.
For lower mass planets, the dissipation in the spiral wake occurs at some distance from the planet \citep{2001ApJ...552..793G,2002ApJ...569..997R,2002ApJ...572..566R} leading to a pair of rings; a similar mechanism may also  lead to a second set of gaps in the dust distribution outside of this first pair; at low-viscosity multiple dust concentrating vortices in the coorbital region, inner, and outer gap edges may form \citep{2017ApJ...843..127D,2017ApJ...850..201B, 2018ApJ...866..110D, 2018ApJ...859..118B,2018ApJ...859..119B}.
The variation of the dust ring and gap features as a result of steady, slow inward migration of the planet has also been described \citep{2018ApJ...866..110D,2019MNRAS.482.3678M}.
In contrast to those cases, in the scenario considered here the features shown here are formed at and adjacent to the planet's radial location, but the planet is able to rapidly migrate inwards and away once the features have been formed.

\section{Discussion}
\label{sec:discussion}

Our simulation results raise a number of important questions, such as how should feedback modified migration ensue when numerical convergence has been achieved, and how should we expect the migration behaviour that we observe to affect the formation of planets and influence the final architectures of planetary systems. These issues are discussed in more detail below.

\subsection{Grid resolution and numerical convergence}
\label{sec:resolution}
Our inviscid disc simulations produce qualitative changes in the migration behaviour of the planets with each change in resolution tested. This is in spite of increasing the number of grid points from 23 to 47 and then 93 cells per scale height, thereby probing the highest resolutions that have been explored to date for these types of simulations. True convergence, by way of meaning an error in the final result which is arbitrarily small, can never be achieved in a discretized non-linear simulation. The next best sense of convergence, that of establishing a series of results in the asymptotic convergence regime, can be aspired to in a numerical model, but our results suggest that at small viscosity values ($\nu < 10^{-7}$) this has not yet been achieved. We interpret this as meaning that as the resolution is increased, the reduced numerical diffusion allows for small scale features such as vortices to emerge more readily in the flow. These can have two effects that combine to produce qualitative changes in the migration of a planet. As discrete structures, they can influence the orbital evolution of the planet directly, introducing varying levels of eccentricity excitation, stochastic behaviour and even sustained periods migration in a particular direction. As structures that introduce dissipation into the flow, they can smooth pressure bumps and gap edges, and hence change the net Lindblad torque acting on a planet. 

It may seem desirable to explore the behaviour at even higher resolutions than considered here, but our highest resolution calculations display structures on scales significantly smaller than the disc scale height, and so a reasonable expectation exists that three dimensional effects may start to have significant influence, particularly in the dynamics of vortices \citep{2009A&A...498....1L}. Thus, at a resolution of 93 zones per scale height, we propose that these simulations may exhaust the usefulness of the vertically averaged inviscid hydrodynamical equations to model inherently three dimensional flows. 

Notwithstanding the uncertainties of the true dynamics of the vortical structures, what we can tell with some certainty is that when the planet mass becomes large enough to create local maxima in the disc radial profile which support Rossby-vortex modes, the influence of these vortices on the migration becomes non-negligible. The resulting orbital evolution becomes stochastic, and the disc radial structure is modified by the presence of these vortices, so the evolution is not well described by slowly evolving one-dimensional models of the gap. The general tendency of the presence of these vortices is to enable continued inward  migration, right into the Type II regime for giant planets. Moreover, the increased tendency for vortices to form at higher resolution leads to sustained vortex-driven planet migration being more pronounced in higher resolution simulations. By physical argument, the increasing role of vortices, 
both with higher resolution and lower physical viscosity is reasonable, and hence we argue that the result that feedback-induced stalling of migration does not occur for planets above the feedback mass is robust.

In contrast to our results, some previous works, such as \citet{2017ApJ...839..100F} and \citet{2018ApJ...859..126F}, have found that planets exceeding the feedback mass efficiently halt their migration in inviscid discs. Although there exist a variety of numerical and physical differences between our calculations and those studies, considering the results here the most likely explanation for the qualitative difference between results is the resolution of those inviscid disc simulations. As we have found, feedback halting can be more effective in inviscid models at low resolution than high. Additionally, we have found in our models that even after a planet has momentarily halted, the migration can resume if sufficiently vigorous RWI occurs at later times. 

One question raised by our calculations remains unanswered, and will require more realistic, high-resolution 3D simulations to be conducted in the future for a definitive conclusion to be reached: do planets with masses between the feedback and thermal masses undergo Type~III migration episodes? The answer to this question will clearly have a strong impact on the final architectures of planetary systems, and also on the observational appearance of protoplanetary discs, as we have shown in Section~\ref{sec:yarfm}. 

\subsection{Implications for planet formation and migration}
\label{sec:formation}
If we combine the results of our previous work on the migration of sub-feedback mass planets in inviscid discs, with and without mid-plane accretion flows \citep{2014MNRAS.444.2031P, 2017MNRAS.472.1565M,2018MNRAS.477.4596M}, with the results obtained in this paper, then we can try to construct a picture of how the migration behaviour obtained in these studies influences the formation and evolution of individual planets and planetary systems. Here, we outline two scenarios. One for a purely wind-driven disc with accretion occurring in thin surface layers and no mid-plane accretion flow, and one for a disc where accretion occurs throughout its full vertical column (an ``advective disc").

\subsubsection{Scenario 1 - A disc with accreting surface layers}
\label{sec:scenario1}
In the discussion below, we assume that the mass contained in the actively accreting layer located near the disc surface is too small to strongly influence a planet's migration, even when the planet emerges from being an embedded object by forming a gap. The robustness of this assumption will need to be tested using future 3D simulations of layered disc models, which go beyond the scope of the current work.
\begin{enumerate}
\item A low-mass planet forms and starts to migrate inwards due to Lindblad torques. As it migrates the planet increases its mass by accreting solids from the disc.
\item Inward migration causes a vortensity contrast to develop between the librating, corotating material and the surrounding disc, so that the dynamical corotation torque acts to slow migration. Continued mass growth of the planet expands the librating region, however, allowing gas to enter which has the same vortensity as the background disc. This lowers the vortensity contrast and weakens the corotation torque so that slow inward migration can continue.
\item At masses of a few Earth-masses, a low-mass hydrogen-helium envelope can settle onto the planet. The planet starts to exceed the feedback mass and a gap forms in the disc. Planet migration starts to stall completely due to feedback effects, and vortices start to form at the pressure bumps located at the gap edges. 
\item In the pebble accretion scenario, the inward drift of pebbles is halted at the gap outer edge, cutting off the feedstock of solids. Growth of the planet's mass is halted. Gas accretion at these low masses is slow even in the absence of solids accretion, unless the envelope opacity is much reduced below interstellar values \citep[e.g.][]{2017MNRAS.470.3206C}.
\item The planet interacts directly with the vortices, leading to stochastic forcing of the planet's orbital evolution. The vortices smooth the gap edges, allowing a net Lindblad torque to drive slow inward migration. If the migration were to proceed as observed in our highest resolution cases, then the migration time-scale at 1 au in an MMSN-like disc is $\sim 50000\ \mathrm{yr}$, independent of the planet mass when it lies between the feedback mass, $M_{\rm F}$  and the thermal mass $M_1$. 
\item The slow inward drift of the planet may be punctuated by periods of stalling due to the formation of one or more coorbital vortices, or episodes of runaway Type~III migration. Slow inward drift may bring the planet into the vicinity of planetesimals or other protoplanets, allowing collisional accretion to occur, although shepherding effects may also intervene to reduce or prevent this. Type~III episodes would reduce shepherding, and may enhance collisional accretion. Growth of the planet to exceed the critical core mass for runaway gas accretion becomes possible, depending on the history of solids accretion. 
\item The stochastic and erratic nature of the migration likely means that the high probability of locking into mean motion resonances in multiplanet systems, as observed in viscous discs for which classical Type~I migration operates \citep{2006A&A...450..833C}, is substantially reduced, modifying the predicted architectures of multiplanet systems compared with previous models.
\item For gas giant planets that form, sustained vortex-driven Type~II inward migration should occur. Unlike our numerical experiments with inviscid discs, gas is continuously supplied to the vicinity of the planet by the surface accretion flows in wind-driven disc models. Hence, pressure bumps can be maintained through continuous tidal truncation of the disc by the planet, and the planet can maintain contact with the outer edge of the gap that drives migration. The migration rate over large scales is controlled by the rate at which gas is supplied to the vicinity of the planet.
\end{enumerate}

\subsubsection{Scenario 2 - An advective disc}
\label{sec:scenario2}
Here we begin by noting that the migration behaviour of a sub-feedback mass planet due to the dynamical corotation torque in an advective disc depends on the planet's Lindblad-torque-driven migration speed versus the gas radial flow speed. \citet{2017MNRAS.472.1565M,2018MNRAS.477.4596M} showed that for a very low mass planet experiencing a small Lindblad torque, the migration would enter into their regime (ii)  i.e.~fast outward migration in regions where there is a sustained accretion torque acting in the disc. For a more massive planet with a larger Lindblad torque the migration would enter their regime (i)  i.e.~inward migration that slows continuously until the planet migration speed locks to the disc flow speed. This results in the following scenario for planet growth and migration in an advective disc.
\begin{enumerate}
\item A low-mass planet forms, and as the gravitational interaction between it and the disc starts to become significant, the migration rate of the planet driven by the Lindblad torque must by definition be smaller than the flow speed of the accreting gas. For reference, at 1 au in a 2 MMSN--like disc, the flow speed of the gas for an accretion rate of $1 \times 10^{-8}$ M$_{\odot}$ yr$^{-1}$ is $8\times10^{-6}\ \mathrm{au\ yr^{-1}}$. Lindblad torques drive planet migration at this speed when the planet mass equals $0.27\ \mathrm{M_{\oplus}}$. At this location of parameter space, the planet is in regime (ii)  of \citet{2017MNRAS.472.1565M,2018MNRAS.477.4596M}, and the planet enters a phase of rapid outward migration.
\item Outward migration can come to a halt for one of a number of reasons: the planet may enter a region where the accretion flow at the disc mid-plane is not present, or differs significantly compared to that at its starting location; the planet may undergo significant mass growth, and hence the vortensity in the corotation region would be modified because the librating region expands and allows background disc material to enter; the planet might interact with and be scattered by another body, losing the librating material that drives the corotation torque.
\item Once the planet jumps out of the outward migration phase, and its mass exceeds the value at which Lindblad torques drive inwards migration at a speed that exceeds the gas inflow speed, the planet will enter regime (i) described by \citet{2017MNRAS.472.1565M,2018MNRAS.477.4596M}. Hence, it will migrate inward and asymptotically lock to the disc flow.
\item As the planet mass grows and exceeds the feedback mass, a gap forms along with vortices at the gap edges. The corotation torque will become sub-dominant and the planet will now migrate at a speed equal to its corresponding vortex-driven migration speed in a disc with no accretion flow, plus the radial flow speed of the accreting gas. The migration speed will be approximately Galilean invariant with respect to the disc accretion flow.
\item The formation of a gap will cut off the flow of pebbles in the pebble accretion scenario, halting the mass growth of the planet. Growth through solids accretion would need to occur through planetesimal accretion or giant impacts with other protoplanets which may be in different migration regimes.
\item The formation of a coorbital vortex with which the planet horseshoe orbits would cause migration to occur at close to the disc radial flow rate. The vortex will be torqued by the global magnetic field and will migrate with the disc flow, likely taking the planet with it should the vortex contain sufficient mass. 
\item Multiple systems of partial gap-opening super-Earths would most likely not experience convergent Type I migration and resonant locking, as they will more or less migrate inwards at the same speed if driven by vortex-smoothing of gap edges . It is perhaps noteworthy that we did not observe any periods of Type III migration in our simulations of planets in advective discs described in Section~\ref{sec:advectivedisc}.
\item We would expect that any giant planets that form would sustain inward vortex-driven Type II migration, with the rate of sustained migration being determined by the rate at which gas is supplied to the vicinity of the planet.
\end{enumerate}

\section{Conclusions}
\label{sec:conclusions}
We have conducted a suite of simulations of planets embedded in inviscid and low-viscosity disc models, with the primary aim of examining the migration of planets whose masses are large enough to modify the local surface density through the non-linear steepening and dissipation of spiral waves. Our numerical set up was designed to be as simple as possible, in order to simplify the analysis of an already complex problem. We employed 2D, globally isothermal discs with power-law surface density profiles,  considered a range of planet masses, disc viscosity values, and evolution in the presence and absence of a laminar accretion flow. Our main conclusions are as follows:
\begin{enumerate}
\item For planets above the feedback mass we find that during early evolution the planet begins to form a gap, and feedback from the resulting asymmetric gap structure causes the planet's migration to slow down. This is in agreement with 1D semi-analytic analyses of the problem that introduced the concept of the inertial mass \citep{1984Icar...60...29H,1989ApJ...347..490W} or the stopping mass \citep{2002ApJ...572..566R}, and with previous 2D hydrodynamical simulations \citep{2009ApJ...690L..52L, 2010ApJ...712..198Y, 2017ApJ...839..100F,2018ApJ...859..126F}.
\item For completely inviscid discs, or for those with very low-viscosity (i.e. dimensionless $\nu < 10^{-7}$), gap formation leads to the formation of pressure bumps that form vortices through the RWI. These have a strong effect on the planet's orbital evolution, exciting the eccentricity, inducing erratic changes in the semimajor axis through stochastic forcing, and smoothing the gap edges so that a net Lindblad torque is maintained that drives inward migration. Vortices may also form in the corotation region, and we speculate that on intermediate to long time-scales these may influence migration by undergoing horseshoe motions with the planet. These effects are inevitably missed by a 1D semi-analytic study of the problem.
\item We have undertaken a resolution study of feedback-modified migration. When changing the resolution from 23 to 47 and 93 grid cells per scale height, we obtain qualitatively different behaviour with each step in resolution. At 47 cells per scale height, vortices are formed and influence the migration through stochastic forcing and gap edge smoothing, and we obtain slow vortex-driven migration punctuated by episodes of runaway Type III migration.  
When moving to 93 cells per scale height, vortices emerge even more readily in the flow, such that short-time-scale stochastic effects are increased. Here, vortex-driven migration is sustained without bursts of Type III migration, with the migration time scales being more or less independent of the planet mass when it lies between the feedback and thermal masses (corresponding to $\sim 3$ and $10\ \mathrm{M_{\oplus}}$ in our simulations). 
Notwithstanding the presence of transient corotational vortex-driven outward migration, the migration time-scale obtained at 1 au in a 2MMSN-style disc is $50000\ \mathrm{yr}$. This is short compared to expected disc life times, but substantially longer than obtained from classical Type~I migration theory.
\item Experiments conducted with varying levels of viscosity show that results consistent with the inviscid models are obtained for $\nu < 10^{-7}$. For intermediate values of $\nu$, gap formation occurs but vortex formation is suppressed, and we observe feedback stalling of migration as predicted from the semi-analytic theory. At higher values of $\nu \sim 10^{-5}$, gap formation is suppressed and we obtained migration rates consistent with classical Type~I migration theory.
\item For gap opening planets above the feedback mass, we find that the background vortensity/surface density gradient influences the migration rate, with migration being noticeably slower for a flatter surface density profile. Although difficult to interpret, this suggests that even in the presence of a gap and in the absence of viscosity, the planet experiences a corotation torque, possibly because of diffusion driven by vortices near the corotation region.
\item Experiments conducted for advective discs that sustain laminar accretion flows in their mid-planes indicate that migration is approximately Galilean invariant with respect to the disc radial flow for planet masses between the feedback and thermal masses. In other words, a planet which migrates at a given speed in an inviscid disc without a mid-plane accretion flow, will migrate at that same speed plus the disc radial flow speed in an advective disc.
\item We observe transient vortex-driven Type~II migration for giant planets embedded in inviscid discs without accretion flows. Here, migration stalls when the planet has moved a distance of the order of the gap width, since vortex-smoothing of the pressure bump sitting outside of the planet is essentially complete once this has occurred, and further migration would break the contact between the planet and the exterior disc.
\item The formation of gaps and pressure bumps blocks the gas drag driven radial migration of small bodies such as pebbles, giving rise to a pebble isolation mass that is significantly smaller than estimates obtained from viscous disc models. Furthermore, the migration of a single planet at large radius in a disc can produce numerous pressure bumps and vortices which efficiently trap mm-sized dust particles. These features are on a scale that allows them to be detected by submillimetre interferometers such as ALMA. Hence, multiple rings and vortex features may indicate the presence of a single migrating planet whose mass is above the local feedback mass, and may not require the presence of multiple bodies in the close vicinity of the observed rings. 
\end{enumerate}

Although the vortex-driven migration of planets found here is very likely a robust phenomenon, the results of our simulations are undoubtedly affected by number of missing physical ingredients from the models. At high resolution, numerous features appear in the flow that are smaller than the disc scale height, suggesting that 3D effects may become important in regulating the flow. The adoption of a globally isothermal disc means that we have a barotropic equation of state, so that the baroclinic driving term  in the vorticity equation, $\nabla \rho \times \nabla P$, is unable to influence the evolution of vorticity. Given the importance of vortices for our results, it seems very likely that relaxing the barotropic assumption will influence the evolution substantially. In turn, a more sophisticated treatment of the thermal evolution through the inclusion of radiation transport is also likely to be important in determining how influential baroclinic effects actually are. We will address these and other issues in future work.

\section*{Acknowledgements}
This research was supported by STFC Consolidated grants awarded to the QMUL Astronomy Unit 2015-2018  ST/M001202/1 and 2017-2020 ST/P000592/1.
This research utilized Queen Mary's Apocrita HPC facility, supported by QMUL Research-IT \citep{apocrita};
and the DiRAC Data Centric system at Durham University, operated by the Institute for Computational Cosmology on behalf of the STFC DiRAC HPC Facility (www.dirac.ac.uk).
This equipment was funded by a BIS National E-infrastructure capital grant ST/K00042X/1, STFC capital grant ST/K00087X/1, DiRAC Operations grant ST/K003267/1 and Durham University. 
DiRAC is part of the National E-Infrastructure.
SJP is supported by a Royal Society University Research Fellowship.
This project has received funding from the European
Union's Horizon 2020 research and innovation programme under grant
agreement no. 748544 (PBL).




\bibliographystyle{mnras}
\bibliography{gapopen_paper} 



\appendix

\section{Grid and boundary conditions}
\label{app:grid}

Here we describe the algorithm used to construct the radially variably spaced grids used in this work.
The grid is evenly spaced in azimuthal angle, and variably space in radius.
Radially it has two regions, with the inner having spacing proportional to radius to give equal-aspect-ratio cells, and 
the outer having a gradually radially degraded resolution, with a transition at radius $r_c$. 
The radial spacing $\delta r$ as a function of radius $r$ is
\begin{align}
\delta r (r) = \delta r_0\, r
\begin{cases}
1 &\text{if } r < r_c\\
\exp\left(\left(\frac{r-rc}{r_0}\right)^2\right)  &\text{if } r \geq r_c
\end{cases}\ .
\end{align}
Here the resolution is set by the base radial grid spacing  $\delta r_0$.
This parameter decreases by a factor of two between each grid specification, 
so for the r2 grid it is $\delta r_0=1.5\times10^{-3}$, for the r4 grid $\delta r_0=7.5\times10^{-4}$, 
and for the r8 grid $\delta r_0=3.75\times10^{-4}$.
To generate a set of grid points grid covering in the interval $[r_i,r_o]$
the grid is built up starting from the inner boundary $s_0=r_i$, successively adding grid points to the set $\{s_0, s_1,...\}$, 
so that the $j+1$th grid point is $s_{j+1}=s_j + \delta r (s_j)$.
When $s_j>r_o$, no more grid points are added,
and the set of grid points is mapped into the set of grid points $r_j$ in the interval $[r_i,r_o]$
by 
\begin{align}
r_j = \frac{s_j-\min(s)}{\max(s)-\min(s)}(r_o-r_i)+r_i\ ,
\end{align}
and this set specifies the radial grid.
The number of evenly spaced grid points used in the azimuthal direction is set to make the cells roughly 
square at $r_0$, so is $2\pi r_0 /\delta r_0$ rounded to the nearest integer.

For the models in an extended domain in Section~\ref{sec:yarfm} the grid parameters are in physical units
$r_0=25\ \text{au}$, $r_c=32\ \text{au}$, $r_i = 5\ \text{au}$, $r_o=60\ \text{au}$ with $\delta r_0=1.5\times10^{-3}\ \text{au}$.

Damping boundary conditions are used as in \citet{2017MNRAS.472.1565M} to minimise wave reflections.
Here, the widths are such that the inner and outer edges have a Keplerian orbital frequency ratio $2/3$ \citep{2016ApJ...826...13B}, 
thus given inner and outer edges of the grid $r_{\rm i}$ and $r_{\rm  o}$ the edges of the damping regions $r_{\rm id}$ and $r_{\rm od}$ are located at:
\begin{align}
r_{\rm id} &= \left(\frac{2}{3} r_{\rm i}^{-3/2}\right)^{-2/3}\, ,\\
r_{\rm od} &= \left(\frac{3}{2} r_{\rm o}^{-3/2}\right)^{-2/3}\, .
\end{align}


\section{Torque density profiles for a disc with a vortensity gradient}
\label{app:vorttorque}

\begin{figure}
\includegraphics[width=\columnwidth]{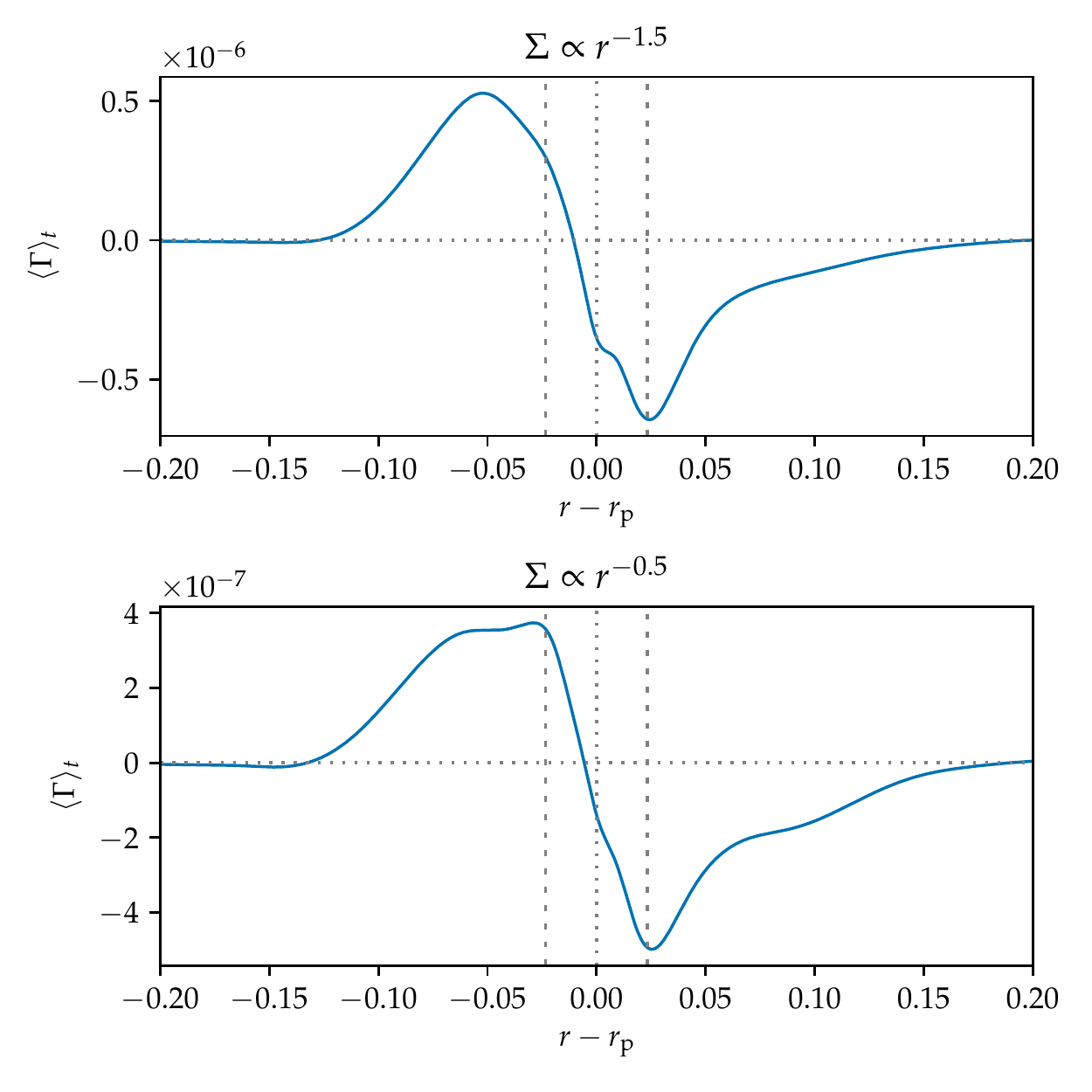}
\caption{Torque density profiles, averaged over the period from 400 to 800 orbits
for the planet with mass ratio $q=3\times 10^{-5}$, inviscid model with 
 no vortensity gradient ($\Sigma\propto r^{-3/2}$),
 and with 
$\Sigma\propto r^{-1/2}$.
The vertical long dashed lines show radial position where a Keplerian flow becomes supersonic with respect to the planet (sonic points). 
}
\label{fig:torquedensity_gap_29_40}
\end{figure}

In section~\ref{sec:vortensitygradient} planets above the feedback mass were observed to migrate more slowly in a disc
with a radial vortensity gradient, than in a disc without.
Figure~\ref{fig:torquedensity_gap_29_40} shows the time-averaged torque densities 
over the period from 400 to 800 orbits in the runs shown in Figure~\ref{fig:gap40}.
In the disc with a vortensity gradient ($\Sigma \propto r^{-0.5}$, lower panel) 
a more positive torque contribution may be seen in particular towards the inner sonic point,
in contrast to the case of a disc without a vortensity gradient ($\Sigma \propto r^{-1.5}$, upper panel).
This behaviour should be the topic of further investigation as it would not necessarily be expected for such a planet with a mass above the feedback mass and near the thermal mass.


\bsp	
\label{lastpage}
\end{document}